\documentclass{aa}

\usepackage{graphicx}
\usepackage{subfigure}
\usepackage{epsfig}
\usepackage{txfonts}
\usepackage{amsgen}
\usepackage{amsfonts}
\usepackage{amssymb}
\usepackage{amsbsy}
\usepackage{natbib}

\begin{document}

	
\title{Tracing the Energetics and Evolution of Dust with {\it Spitzer}: \\
a Chapter in the History of the Eagle Nebula}

\author{
  N. Flagey\inst{1,2}
  \and
  F. Boulanger\inst{2}
  \and
  A. Noriega-Crespo\inst{1}
  \and
  R. Paladini\inst{1}
  \and
  T. Montmerle\inst{3,4}
  \and
  S.J. Carey\inst{1}
  \and
  M. Gagn\'e\inst{5}
  \and
  S. Shenoy\inst{1,6}
}
	
\offprints{N. Flagey}
	
\institute{
  Spitzer Science Center, California Institute of Technology, 1200 East California Boulevard, MC 220-6, Pasadena, CA 91125, USA
  \email{nflagey@ipac.caltech.edu}
  \and
  Institut d'Astrophysique Spatiale, Universit\'e Paris Sud, B\^at. 121, 91405 Orsay Cedex, France
 \and
  Institut de Plan\'{e}tologie et d'Astrophysique de Grenoble, BP53, 38041 Grenoble Cedex 9, France, 
  \and
  Institut d'Astrophysique de Paris, 98bis, Bd Arago, 75014 Paris, France
  \and
  Department of Geology and Astronomy, West Chester University, West Chester, PA 19383, USA
  \and
  Space Science Division, Mail Stop 245-6, NASA Ames Research Center, Moffett Field, CA 94035, USA
}
	
\date{Received ; accepted }

	
\abstract
{The Spitzer GLIMPSE and MIPSGAL surveys have revealed a wealth of details of the Galactic plane in the infrared (IR). We use these surveys to study the energetics and dust properties of the Eagle Nebula (M16), one of the best known SFR.}
{We present MIPSGAL observations of M16 at 24 and 70 $\mu$m and combine them with previous IR data. The mid-IR image shows a shell inside the well-known molecular borders of the nebula. The morphologies at 24 and 70 $\mu$m are quite different, and its color ratio is unusually warm. The far-IR image resembles the one at 8 $\mu$m that enhances the structure of the molecular cloud and the Pillars of creation. We use this set of data to analyze the dust energetics and properties within this template for Galactic SFR.}
{We measure IR SEDs across the entire nebula, both within the shell and the PDRs. We use the DUSTEM model to fit these SEDs and constrain dust temperature, dust size distribution, and interstellar radiation field (ISRF) intensity relative to that provided by the star cluster NGC6611.}
{Within the PDRs, the dust temperature, the dust size distribution, and the ISRF intensity are in agreement with expectations. Within the shell, the dust is hotter ($\sim 70\ \rm{K}$) and an ISRF larger than that provided by NGC6611 is required. We quantify two solutions to this problem. (1) The size distribution of the dust in the shell is not that of interstellar dust. (2) The dust emission arises from a hot ($\sim 10^6 \,$K) plasma where both UV and collisions with electrons contribute to the heating.}
{We suggest two interpretations for the M16’s inner shell. (1) The shell matter is supplied by photo-evaporative flows arising from dense gas exposed to ionized radiation. The flows renew the shell matter as it is pushed out by the pressure from stellar winds. Within this scenario, we conclude that massive star forming regions such as M16 have a major impact on the carbon dust size distribution. The grinding of the carbon dust could result from shattering in grain-grain collisions within shocks driven by the dynamical interaction between the stellar winds and the shell. (2) We also consider a more speculative scenario where the shell would be a supernova remnant. We would be witnessing a specific time in the evolution of the remnant where the plasma pressure and temperature would be such that the remnant cools through dust emission.}
	

\maketitle


\section{Introduction}
\label{lab:intro}

The Eagle Nebula (\object{M16}) is a nearby \citep[$\rm d= 2.0 \pm 0.1$~kpc,][]{Hillenbrand1993} massive star forming region made a sky icon by the publication of spectacular Hubble Space Telescope (HST) images of the ionized gas emission \citep{Hester1996}. As one of the nearest star forming region and one of the most observed across the electromagnetic spectrum, the Eagle Nebula is a reference source. The nebula cavity is carved into the molecular cloud by a cluster of 22 ionizing stars earlier than B3 \citep{Dufton2006} and with an estimated age of $1 - 3\times 10^6$ yrs \citep{Hillenbrand1993, Dufton2006, Martayan2008}.

The mid-IR images of M16 either from the Infrared Space Observatory Camera \citep[ISOCAM][]{Cesarsky1996} at 8 and 15 $\mu$m \citep{Pilbratt1998, Omont2003} or based on the combined {\it Spitzer} observations using IRAC 8 $\mu$m \citep{Fazio2004} and MIPS 24 $\mu$m \citep{Rieke2004}, show a shell-like emission at 15 and 24 $\mu$m that fills the nebula cavity \citep{Flagey2009b}, as delineated by the shorter IR wavelengths and the extent of the H$_\alpha$ emission. The shell stands out in the ISO 15 $\mu$m and MIPS 24 $\mu$m images, while the Nebula pillars, and the outer rim of the nebula are the strongest emission features at 8 $\mu$m. Based on some spectroscopic evidence \citep{Urquhart2003}, we know that the mid-IR shell emission arises from dust with only a minor contribution from ionized gas lines to the broadband emission.

M16 is not alone in this respect. There are other large, partially symmetrical and rich HII regions (in terms of their OB stellar content) that display a similar mid-IR color stratification: the Rosette Nebula \citep{Kraemer2003}, the Trifid Nebula \citep{Lefloch1999, Rho2006}, and M17 \citep{Povich2007}. Furthermore, the multi-wavelength observations of the HII regions in the Galactic Plane, using the {\it Spitzer} GLIMPSE and MIPSGAL Legacy surveys \citep{Churchwell2009, Carey2009} display overall a wide variety of complex morphologies, and show many ``bubble''-like objects with a similar color stratification as M16 \citep{Watson2008, Watson2009}, although they are smaller and driven by one or a few OB stars.

\textbf{What are these Spitzer images of massive star forming regions teaching us about dust and the interaction of the stars with their environment? The IRAC and the MIPS $24\mu$m camera are imaging the emission from PAHs and Very Small Grains (VSGs). A first key to the interpretation of Spitzer images is the change in abundance and excitation of these small dust particles from molecular to ionized gas. Observations of nearby molecular clouds illuminated by O stars, where observations separate the H~II photo-ionized gas layer from the neutral Photo-Dissociation Region (PDR) show that the PAH bands, which are a characteristic of PDR mid-IR emission spectra, are strikingly absent from that of the H~II layer \citep[e.g the Orion Bar and the M17SW interface, ][]{Giard1994b, Cesarsky1996b, Povich2007}. PAHs are quickly destroyed when matters flows across the ionization front. Several destruction mechanisms have been proposed: chemisputtering by protons and photo-thermo dissociation and/or Coulomb explosion associated with absorption of high energy photons. Much less is known about the evolution of VSGs. The mid-IR shells may reflect dust processing by hard photons and shocks that impact the fraction of the dust mass in VSGs, but this possibility has yet to be constrained by modeling of the dust emission.}

\textbf{The evolutionary stage of the massive forming regions is a second key to the interpretation of the Spitzer images. The mid-IR shells do not fit the classical view of the evolution of HII regions where the matter is swept away by the simultaneous effect of the ionization, stellar winds, and radiation pressure from their central OB stars \citep[e.g.][]{TenorioTagle1982, Beltrametti1982, Rozyczka1985}. In this scenario, the HII regions are ``hollow''. One interesting possibility is that gas photo-evaporating from dense condensations exposed to ionized radiation, creates a gas mass input within the cavity sufficient to balance the outward flow of matter. Are the shells reflecting such a mass input? To show that this is a plausible interpretation, one must quantify the mass input, as well as the dust properties and excitation conditions, required to match the shells brightness and its distinct mid-IR colors.}

\textbf{So far most of the studies on the mid-IR properties of these HII regions and smaller bubbles have been phenomenological and looking into the spatial distribution of the different emission components and not their physics. A small bubble where a more quantitative analyis has been carried out is G28.82-0.23 \citep{Everett2010}. G28.82-0.23 ({\it aka} N49) is nearly spherically symmetric, excited by a single O5V star, which has a thick 8 $\mu$m shell surrounding at 24 $\mu$m a diffuse bubble \citep[see e.g.][Fig. 7]{Watson2008}. \citet{Everett2010} proposed a model where the mid-IR emission of G28.82-0.23 arises from dust entrained by the stellar wind. This interpretation involves a hot ($> 10 ^6$ K), high pressure plama ($p/k \sim 10^9\ \rm{K.cm^{-3}}$) where dust lifetime is shorter than the expansion timescale. It requires that dust is constantly replenished by photo-evaporation of high density (10$^5\ \rm{cm^{-3}}$) dusty gas cloudlets that have been overrun by the expanding nebula. Collisional excitation by hot electrons contribute significantly to the heating of dust. Infrared dust emission is the dominant cooling channel of the dusty wind, which reduces the energy available for wind-driven expansion. It seems to us that this specific model does not offer a general framework to interpret observations of larger HII regions, where one observes a similar 8 and 24 $\mu$m color stratification.}

The motivation of this paper is to study the nature of mid-IR shells in massive star forming regions using the Eagle Nebula as a template source. The detailed data available on this nearby nebula allow us to perform a quantitative modeling of the dust heating by UV radiation and, possibly, by collisions in a hot plasma. We quantify the dust emission in terms of dust physics, before discussing possible interpretations within an evolutionary scenario of the Eagle Nebula as a massive star forming region. In section \ref{lab:obs}, we present the Spitzer imaging observations of the Eagle Nebula from the MIPSGAL Galactic plane survey. Section \ref{lab:obsres} describes the morphology of M16 based on IR photometric and spectroscopic observations. We measure the spectral energy distribution (SEDs) across the entire nebula combining data from the ISO, MSX and Spitzer space missions.\textbf{ In sections \ref{lab:uvheating} and \ref{lab:collheating}, we present exhaustive modeling of the dust properties. We first model the dust SEDs with UV heating only, and this sets constraints on the radiation field intensity and dust size distribution. Then we consider the possibility that the shell emission arises from a hot plasma where dust would be heated by collisions with electrons. The reader not interested in the details of the modeling can skip sections \ref{lab:uvheating} and \ref{lab:collheating}. In section \ref{lab:nature}, we propose two scenarios of the present evolutionary state of the Eagle Nebula, which could account for the mid-IR shell and fit within present observational constraints.} The paper results are summarized in section \ref{lab:ccl}.

\begin{figure*}[!ht]
	\centering
	\includegraphics[width=\linewidth]{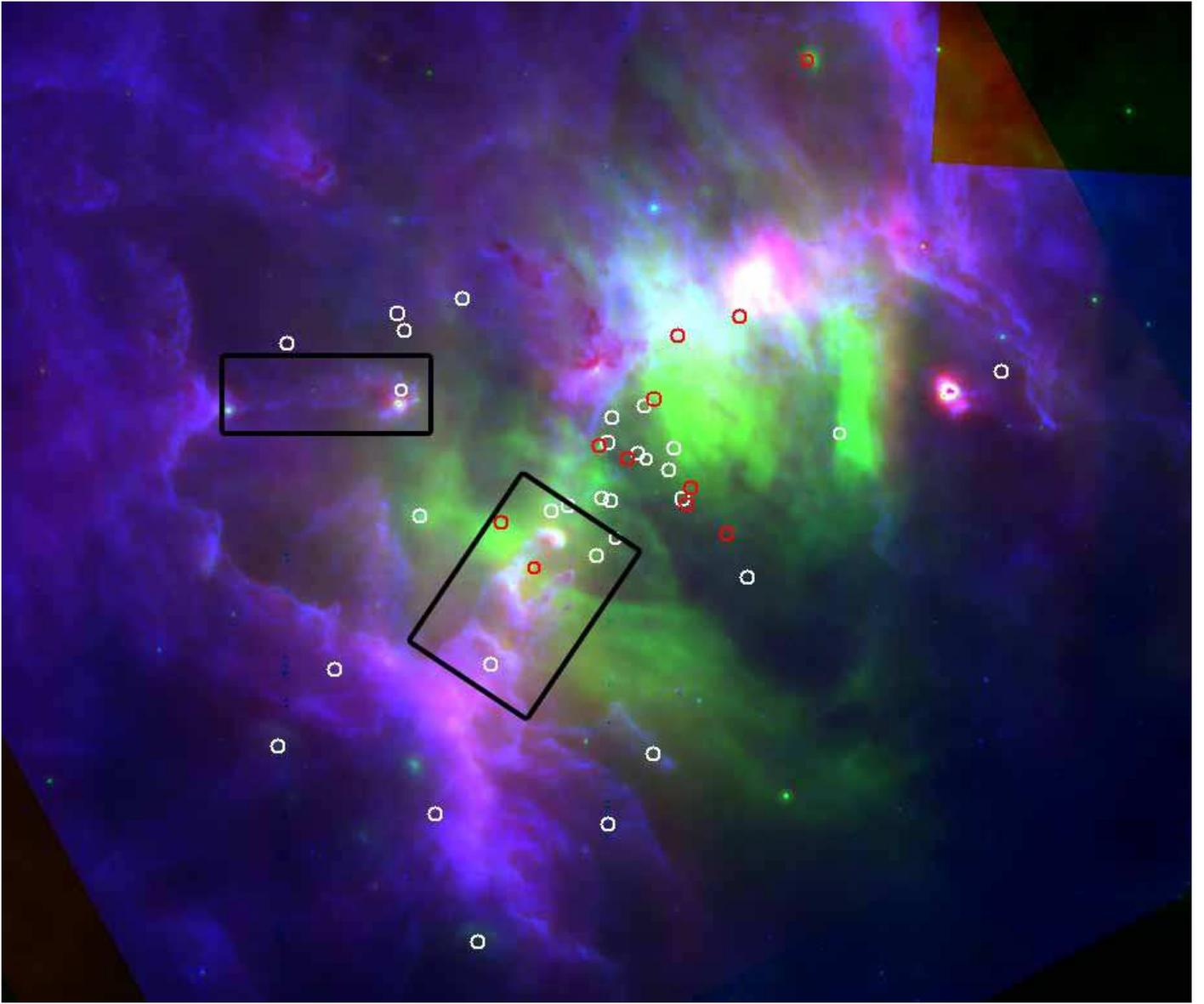}
	\caption{Composite Spitzer color image combining the IRAC 5.8 $\mu$m (blue) bands with MIPS 24 (green) and 70 $\mu$m (red). The FOV is $\sim 30$\arcmin, N is up and E is left. The two black boxes outline the Pillars of Creation, which raise from the bottom to the center, pointing slightly to the West and the Spire, on the East, almost pointing straight toward the West. The position and spectral type of the most massive stars of NGC6611 is overplot: O stars are in red, B stars are in white.}
	\label{fig:m16}
\end{figure*}

\section{Observations}
\label{lab:obs}

The Eagle Nebula has recently been observed by the \textit{Spitzer Space Telescope} as part of the GLIMPSE \citep[program \#00146,][]{Benjamin2003} and MIPSGAL \citep[program \#205976,][]{Carey2009} inner Galaxy surveys. The GLIMPSE survey has made use of the Infrared Array Camera \citep[IRAC,][]{Fazio2004}, while MIPSGAL has been realized with the Multiband Imaging Photometer for Spitzer \citep[MIPS,][]{Rieke2004}. In both cases we have used their enhanced products \citep{Squires2005}. The MIPSGAL 24 $\mu$m data has been complemented with archival observations (Spitzer program \#20726) and reprocessed using the standard Spitzer Post-Basic Calibrated Data tools\footnote{http://ssc.spitzer.caltech.edu/postbcd/}. A three-color image combining IRAC and MIPS data is shown on figure \ref{fig:m16}.

Most of the data processing performed on the MIPSGAL 24 $\mu$m observations is described in \citet{Mizuno2008} and \citet{Carey2009}. At 70 $\mu$m, Spitzer detectors are Ge:Ga photoconductors. When observing bright, structured emission, like the one in the Eagle Nebula, such detectors show significant variations in responsivity, which manifest themselves as visible stripes in the final images, and result in photometric errors of several tens of percent. This effect has required an offline reprocessing of the data, with tools specifically designed to, at the same time, reconstruct the history-dependent responsivity variations of the detectors and mitigate the associated stripes. The photometric uncertainty of extended emission is lowered from about 50\% on the brightest features down to about 15\% on the enhanced MIPS 70 $\mu$m data. The specific pipeline developed for the MIPSGAL 70 $\mu$m observations will be detailed in \citet{Paladini2011}.

\begin{figure}[]
\centering
\subfigure[] 
	{\label{fig:cvf_spec_c}
	\includegraphics[angle=90,width=.45\linewidth]{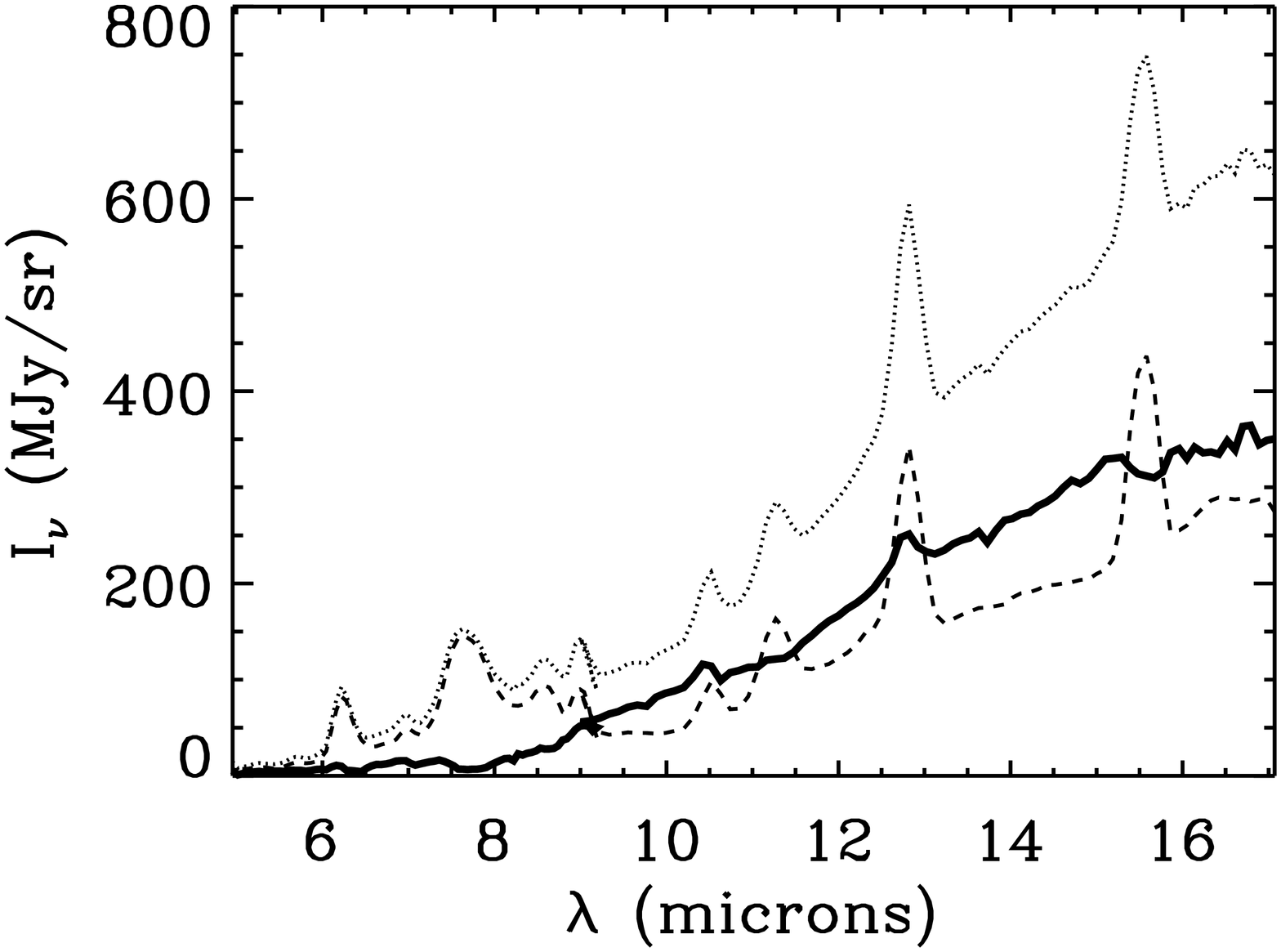}}
\hspace{.5cm}
\subfigure[] 
	{\label{fig:cvf_spec_a}
	\includegraphics[angle=90,width=.45\linewidth]{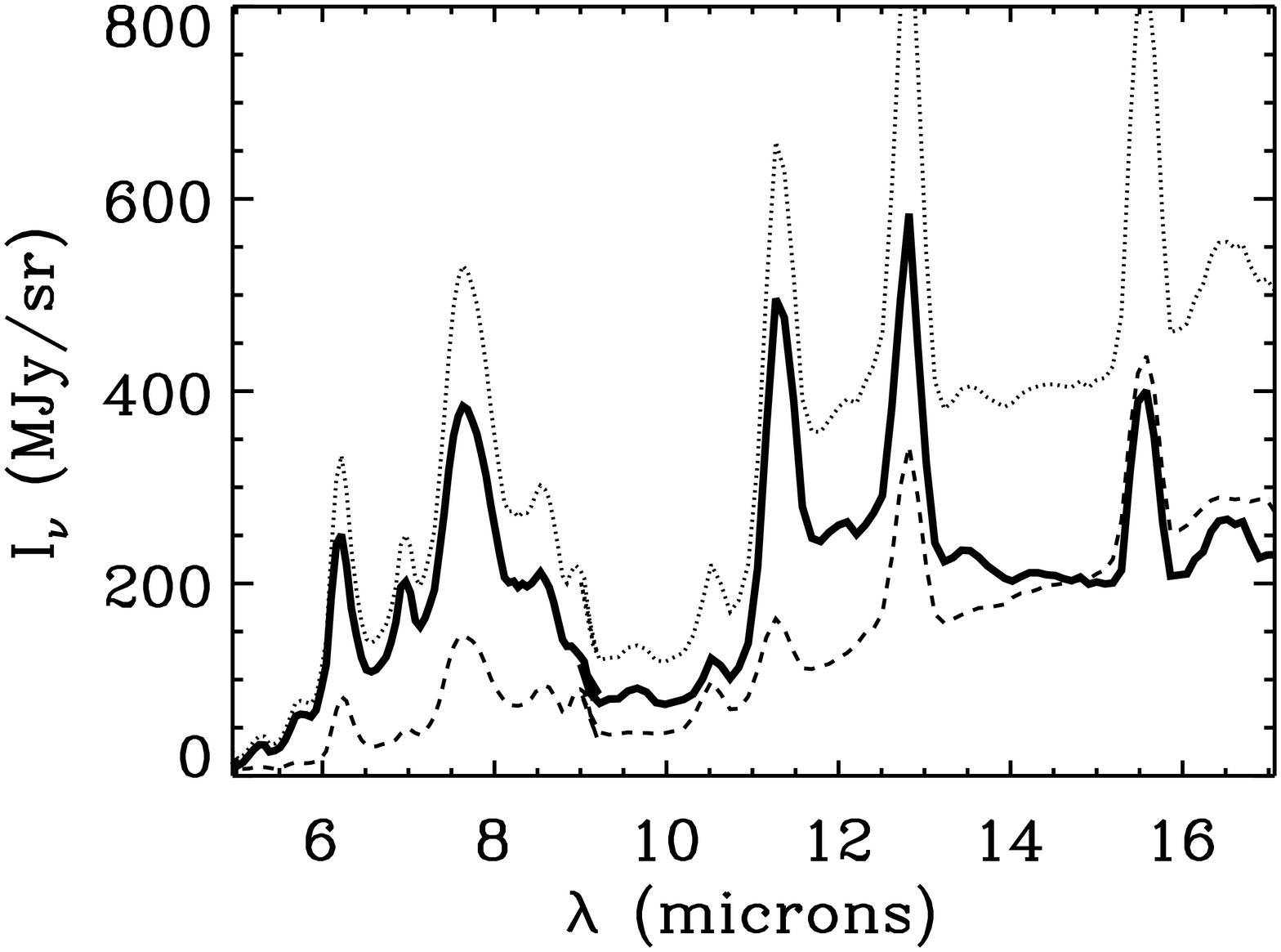}}
\\
\subfigure[] 
	{\label{fig:cvf_spec_b}
	\includegraphics[angle=90,width=.45\linewidth]{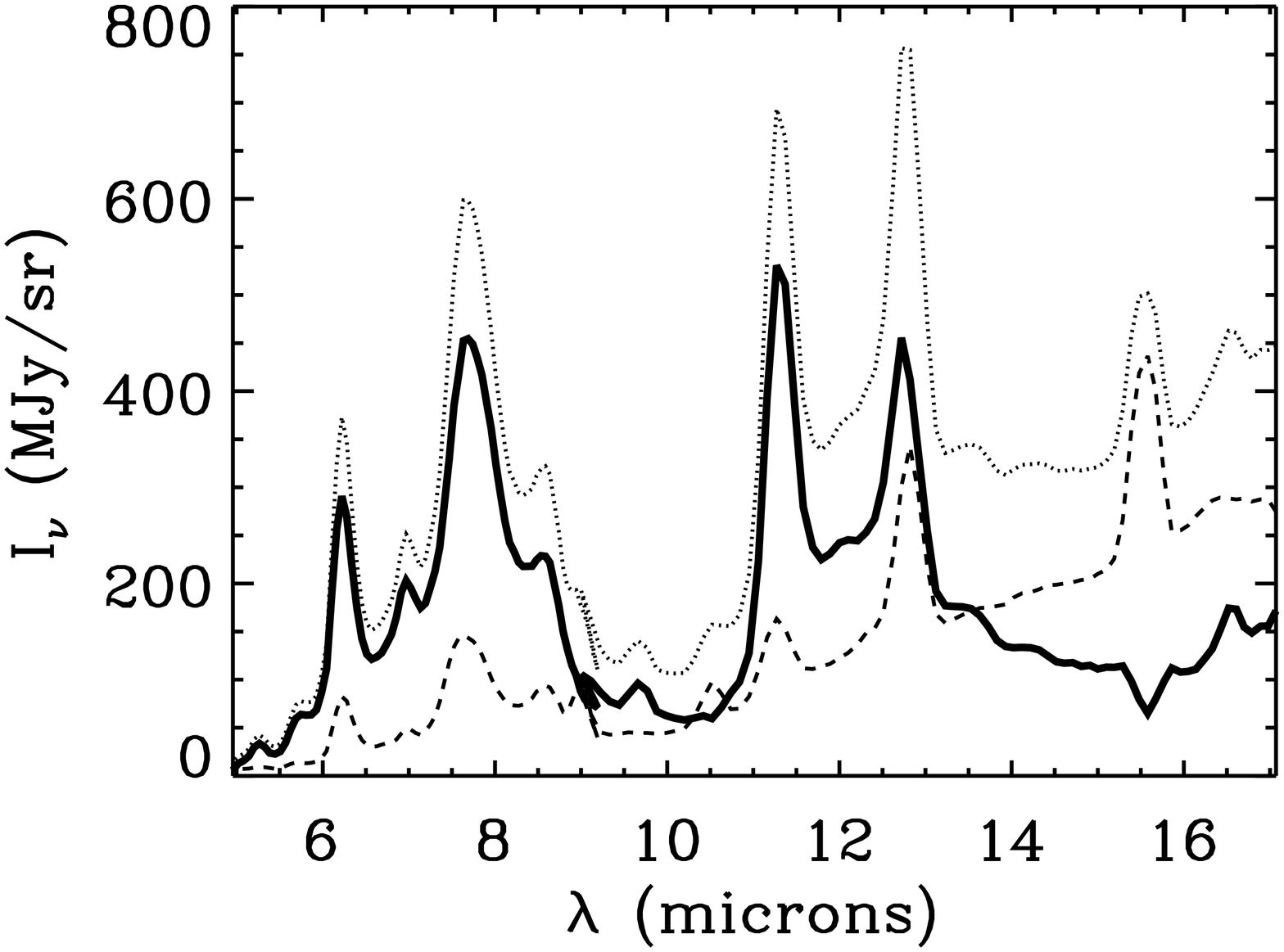}}
\hspace{.5cm}
\subfigure[] 
	{\label{fig:cvf}
	\includegraphics[width=.45\linewidth]{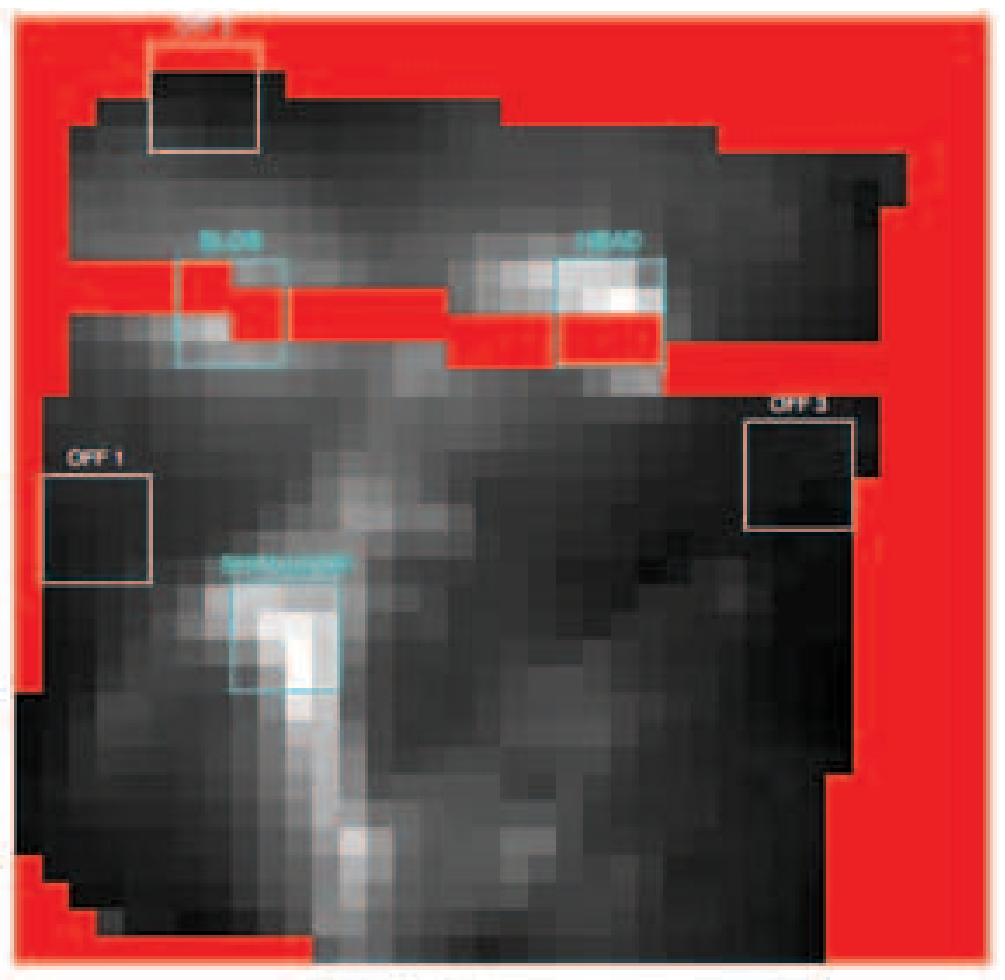}}
\caption{ISOCAM/CVF mean spectra observed (a) on Pilbratt's blob, (b) at the tip of the main Pillar of Creation and (c) within the Pillar of Creations. Dotted lines are ON spectra, dashed lines are OFF spectra, thick solid lines are ON-OFF spectra. OFF and ON positions are shown on the ISOCAM/CVF 3\arcmin\ by 3\arcmin\ field of view, here at the wavelength of 12 $\mu$m. North is up, East to the left.}
\label{fig:cvf_spec}
\end{figure}

We complete the Spitzer observations of M16 with previous IR survey from MSX and observations from ISO, both photometric and spectroscopic. The ISOCAM/CVF spectra have already been presented by \citet{Urquhart2003}. A slice of the ISOCAM/CVF spectroscopic cube is shown on Fig.~\ref{fig:cvf}.

\section{Observational results}
\label{lab:obsres}

We use the many IR observations available to create a portrait of the nebula from NIR to FIR wavelengths. We then perform aperture measurements on both the broad band images and spectroscopic observations in order to get characteristic spectral energy distributions (SEDs) and spectra of the Eagle Nebula. We focus our comments on the two main features of the nebula: the PDRs and the inner shell.

\subsection{Images}
\label{lab:ima}

The three-color image of Fig.~\ref{fig:m16} clearly highlights differences between intermediate wavelengths on the one side (MIPS24 in green) and the shorter and longer wavelengths on the other side (IRAC8 in red and MIPS70 in blue). The whole molecular cloud appears in purple while the inner shell is green.

\begin{itemize}
\item At wavelengths shorter than $\sim10\ \mu$m, IRAC, MSX and ISO observations show the molecular cloud surface heated by the cluster UV radiation. The Pillars of Creation, the Spire (see Fig.~\ref{fig:m16} to identify these structures) and less contrasted emission extend towards the cluster from the North and the East. To the NW and the SE, the rim of an outer shell can be identified. It corresponds to the edge of the Eagle Nebula as seen in H$\alpha$.
\item At intermediate wavelengths, between $\sim12$ and 24 $\mu$m, MSX, ISO and MIPS observations exhibit a significantly distinct morphology, with a shell filling the inside cavity in between the Pillars and the edges of the molecular cloud seen at shorter wavelengths. The shell extends over $\sim12\arcmin$ in the NW-SE direction towards the pillars and further out to the SW where there is no emission either at shorter or longer wavelength. There are some bright features within the shell, some of which have already been identified \citep[e.g. Pilbratt's blob, to the East of the main Pillar of Creation][]{Pilbratt1998}. The lack of far-infrared observations prevented previous authors to conclude anything specific on the nature of this shell.
\item At longer wavelength, the MIPS 70 $\mu$m observations are very similar to those at shorter wavelengths and mainly show the molecular cloud surface.
The diffuse emission within the inside cavity is visible but not as bright as at intermediate wavelengths. The lower angular resolution of these observations does not allow us to make more detailed comments at this point.
\end{itemize}

The IR morphology of the Eagle Nebula is common among other star forming regions. \citet{Churchwell2006} have listed many such ``bubbles'' across the entire GLIMPSE Galactic plane survey with IRAC. Combining GLIMPSE and MIPSGAL 24 $\mu$m surveys reveals an inner shell for most of these regions\footnote{http://www.spitzer.caltech.edu/Media/releases/ssc2008-11/ssc2008-11a.shtml}.

\subsection{SEDs measurements}
\label{lab:seds}

We perform ON-OFF aperture measurements to get both spectroscopic and photometric SEDs. As shown on figure \ref{fig:cvf} there is a band of unavailable pixels on the ISOCAM/CVF observations. This band goes exactly through interesting and contrasted features like the tip of the main Pillar and Pilbratt's blob. Rather than linearly interpolate the missing pixels like it has been done previously on ISOCAM/CVF data \citep[e.g.][]{Urquhart2003}, we use these data as is. We present and interpret spectroscopic and photometric measurements separately.

\subsubsection{Spectroscopic measurements}
\label{lab:specsed}

We compute average spectra on multiple positions within the Pillars of Creation area covered by the ISOCAM/CVF data. We use square boxes of 4x4 pixels (24x24\arcsec on ISOCAM/CVF 6\arcsec pixel field of view) to estimate the mean brightness of several features. We use this method for both ``ON'' and ``OFF'' positions. We combine three different OFF positions to build a unique OFF spectrum. The resulting ON-OFF spectra are shown on figure \ref{fig:cvf_spec} for two positions within the main Pillar of Creation and one on Pilbratt's blob. These three positions, marked on figure \ref{fig:cvf}, correspond respectively to spectra D, B and A of figure 2 from \citet{Urquhart2003}. One of our OFF positions is close to their spectrum C. As a consequence, our results are similar to theirs:

\begin{itemize}
\item The spectra of the Pillars of Creation (see Fig.~\ref{fig:cvf_spec_a} and \ref{fig:cvf_spec_b}) exhibit the characteristics of PDRs spectra with strong PAH features and gas lines. They also present the Si absorption feature around 10 $\mu$m. There are some variations between the two positions, mainly regarding PAHs features and gas lines strength, which traces variations in the excitation conditions between these two positions within the column of gas and dust.
\item The spectrum of Pilbratt's blob (see Fig.~\ref{fig:cvf_spec_c}) exhibits a strong continuum with very weak gas lines and PAHs bands. We thus assume, as a first approximation, that the MIPS 24 $\mu$m shell is dust continuum dominated.
\item The OFF position has a spectrum with a weaker continuum than the blob but stronger than the Pillars. It also has much weaker lines and features than within the gaseous and dusty columns.
\end{itemize}

\subsubsection{Photometric measurements}
\label{lab:photsed}

\begin{figure}[!t]
\centering
\subfigure[] 
	{\label{fig:pillar_box}
	\includegraphics[width=.7\linewidth]{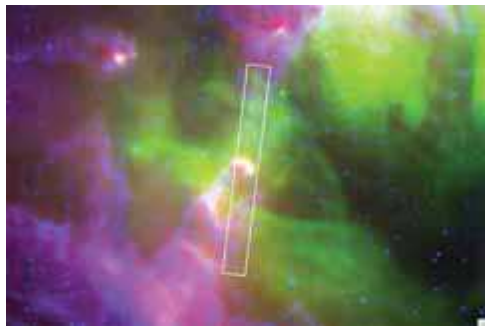}}
\subfigure[] 
	{\label{fig:cut_plot_pillar}
	\includegraphics[angle=90, width=.7\linewidth]{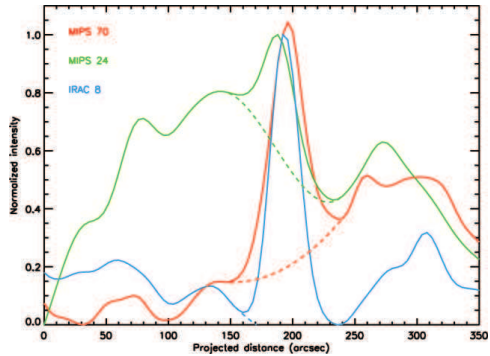}}
\caption{(a) Three color image as in figure \ref{fig:m16} with the region along which the profiles are measured for the main Pillar of Creation. (b) Normalized infrared emission profiles (MIPS70 in red, MIPS24 in green and IRAC8 in blue, solid lines) and interpolations performed to measure the fluxes of the structure (dashed lines).}
\label{fig:cut_n_fit_pillar}
\end{figure}

\begin{figure}[!t]
\centering
\subfigure[] 
	{\label{fig:fairy_box}
	\includegraphics[width=.7\linewidth]{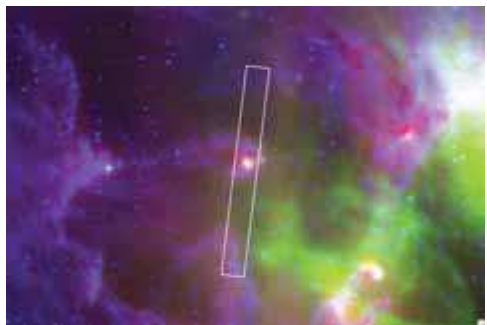}}
\subfigure[] 
	{\label{fig:cut_plot_fairy}
	\includegraphics[angle=90, width=.7\linewidth]{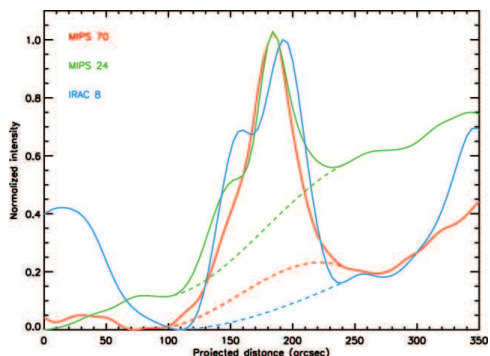}}
\caption{Same as figure \ref{fig:cut_n_fit_blob} for the position of the ``Spire''.}
\label{fig:cut_n_fit_fairy}
\end{figure}

\begin{figure}[!t]
\centering
\subfigure[] 
	{\label{fig:pdr_box}
	\includegraphics[width=.7\linewidth]{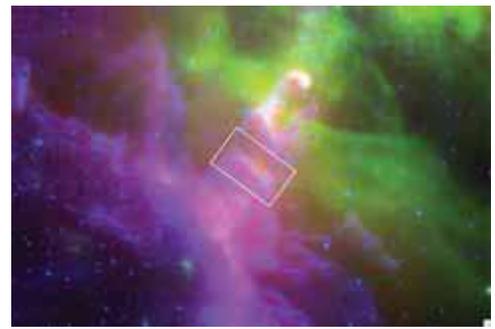}}
\subfigure[] 
	{\label{fig:cut_plot_pdr}
	\includegraphics[angle=90, width=.7\linewidth]{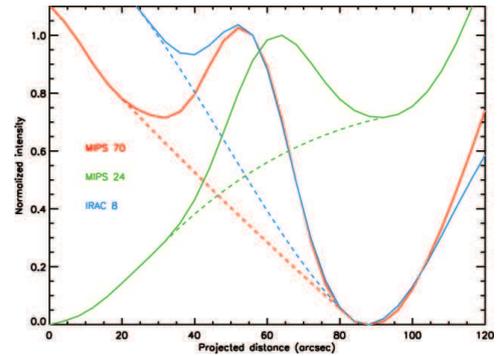}}
\caption{Same as figure \ref{fig:cut_n_fit_blob} for the position of the ``Shoulder''.}
\label{fig:cut_n_fit_pdr}
\end{figure}

\begin{figure}[!t]
\centering
\subfigure[] 
	{\label{fig:blob_box}
	\includegraphics[width=.7\linewidth]{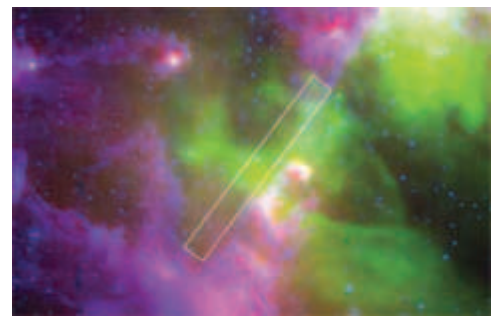}}
\subfigure[] 
	{\label{fig:blob_cut_plot}
	\includegraphics[angle=90, width=.7\linewidth]{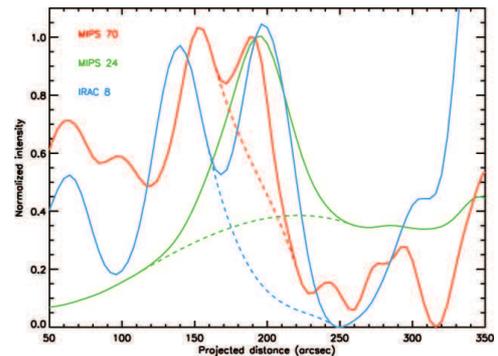}}
\caption{Same as figure \ref{fig:cut_n_fit_blob} for the position of the ``Blob''.}
\label{fig:cut_n_fit_blob}
\end{figure}

\begin{figure}[!t]
\centering
\subfigure[] 
	{\label{fig:shellborder_box}
	\includegraphics[width=.7\linewidth]{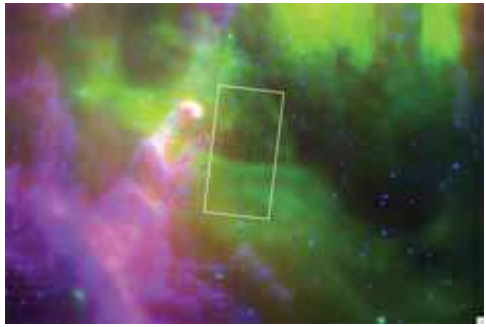}}
\subfigure[] 
	{\label{fig:cut_plot_shellborder}
	\includegraphics[angle=90, width=.7\linewidth]{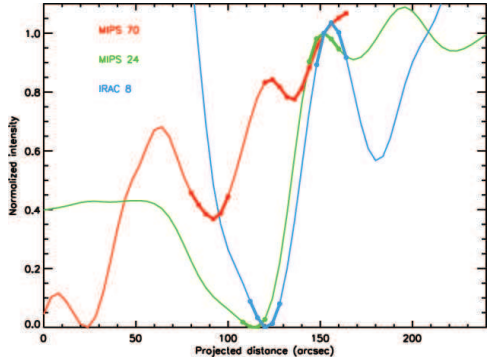}}
\caption{Same as figure \ref{fig:cut_n_fit_blob} for the position of the ``Shell Border''. The darker sections of the profiles show the top and bottom of the ``jump'' used to measure the fluxes at each wavelength.}
\label{fig:cut_n_fit_shellborder}
\end{figure}

\begin{figure}[!t]
\centering
\subfigure[] 
	{\label{fig:backshell_box}
	\includegraphics[width=.7\linewidth]{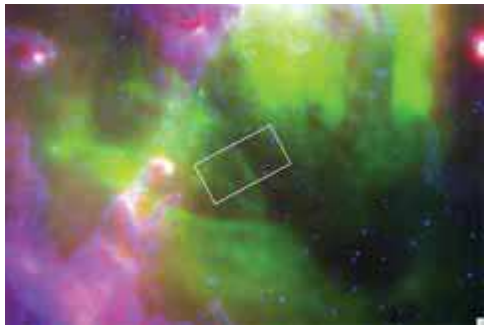}}
\subfigure[] 
	{\label{fig:cut_plot_backshell}
	\includegraphics[angle=90, width=.7\linewidth]{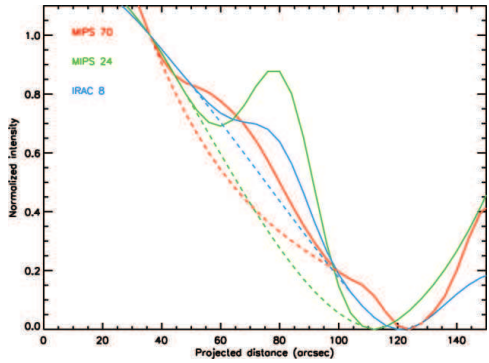}}
\caption{Same as figure \ref{fig:cut_n_fit_blob} for the position of the ``Reverse Shell''.}
\label{fig:cut_n_fit_backshell}
\end{figure}

\begin{figure}[!t]
\centering
\subfigure[] 
	{\label{fig:fila_box}
	\includegraphics[width=.7\linewidth]{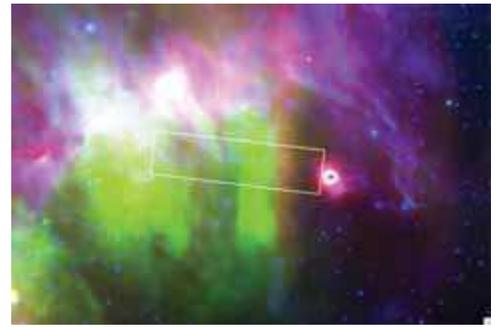}}
\subfigure[] 
	{\label{fig:cut_plot_fila}
	\includegraphics[angle=90, width=.7\linewidth]{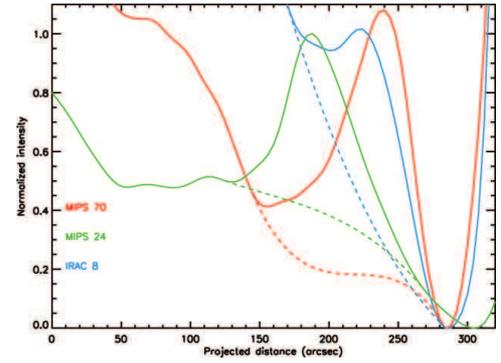}}
\caption{Same as figure \ref{fig:cut_n_fit_blob} for the position of the ``Filament''.}
\label{fig:cut_n_fit_fila}
\end{figure}

\begin{figure}[!t]
\centering
\subfigure[] 
	{\label{fig:diff_box}
	\includegraphics[width=.7\linewidth]{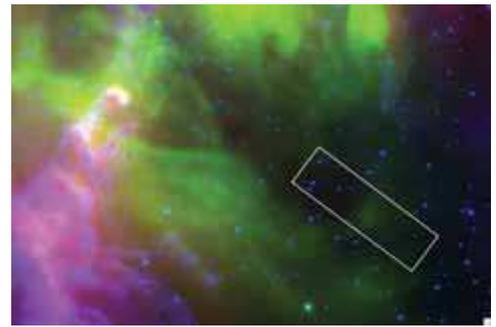}}
\subfigure[] 
	{\label{fig:cut_plot_diff}
	\includegraphics[angle=90, width=.7\linewidth]{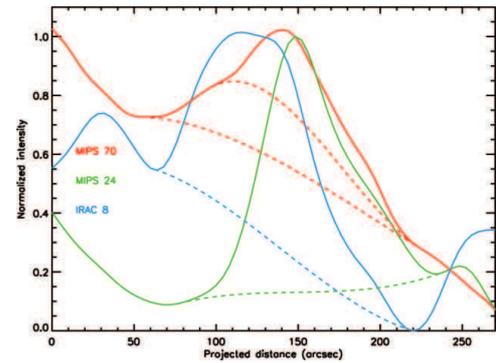}}
\caption{Same as figure \ref{fig:cut_n_fit_blob} for the position of the ``Diffuse''.}
\label{fig:cut_n_fit_diff}
\end{figure}

We combine IR observations of the Eagle Nebula from three different observatory: MSX, ISO and Spitzer. Therefore, we first lower the spatial resolution of each observations to that of MSX data (20\arcsec). Then, as we did with the spectroscopic measurements, we pick up several interesting and contrasted features within the nebula. We name them as follows. The ``PDR'' group of features contains the tip of the main Pillar of Creation (``Pillar'', also known as Column I, with an embedded source at its tip, see Fig.~\ref{fig:pillar_box}), the tip of the Spire (``Spire'', also known as Column IV, with an embedded source at its tip, see Fig.~\ref{fig:fairy_box}) and a PDR within the main Pillar of Creation (``Shoulder'', see Fig.~\ref{fig:pdr_box}). The ``Shell'' group of features contains Pilbratt's blob (``Blob'', see Fig.~\ref{fig:blob_box}), the contrasted border of the main shell (``Shell border'', see Fig.~\ref{fig:shellborder_box}), a diffuse shell that extends towards the opposite direction (``Reverse shell'', see Fig.~\ref{fig:backshell_box}), a bright filament on the North-West side of the nebula(``Filament'', see Fig.~\ref{fig:fila_box}) and some more diffuse emission on the South-West side of the nebula (``Diffuse'', see Fig.~\ref{fig:diff_box}). For each structure, the main difficulty of the measurement is to properly estimate the background emission behind each of them. This is particularly true for the MIPS 70 $\mu$m images.

We illustrate our method on the example of Pilbratt's blob but it is mainly valid for the whole set of structures. We first select a rectangular area that encompasses the blob, as shown on Fig.~\ref{fig:blob_box}. We choose the orientation of the selected area in such a way we avoid to select other neighboring contrasted features (e.g. the Pillars of Creation). We then compute the mean profile of the blob and its surrounding by averaging all the pixels along the short axis. The resulting normalized profiles for Pilbratt's blob are shown on Fig.~\ref{fig:blob_cut_plot} for several wavelengths. The profiles for the other features are shown on Fig.~\ref{fig:cut_plot_shellborder} to \ref{fig:cut_plot_diff}.

We then measure the mid to far-IR SED of each structure. We adapt the method as a function of the profile shape. For the structures that present a peak of emission at every wavelength (e.g. Pilbratt's blob, Spire), we estimate the background through a spline interpolation of the profile on both sides of the peak (see Fig.~\ref{fig:cut_plot_pdr}). The flux of the structure is thus given by the integration of the background subtracted profile over the size of the structure. The actual size over which we integrates the flux may slightly vary from one channel to another. The uncertainty on each measurement is given by the range of background values as estimated by the spline interpolation. For the other structures, where the profiles exhibit a ``jump'' (case of the shell border, see Fig.~\ref{fig:cut_plot_shellborder}), we estimate the height of the ``jump'' at each wavelength by measuring the difference of the surface brightness between the top and bottom of the ``jump''. The uncertainty on each measurement is given by the standard deviation of the surface brightness at the top and the bottom of the ``jump''.

\begin{figure}[]
	\centering
	\includegraphics[angle=90, width=.9\linewidth]{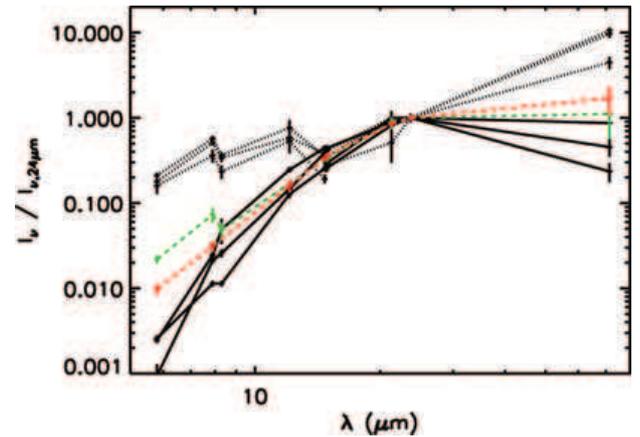}
	\caption{Comparison of the structures SED. Solid lines: structures within the shell. Dotted lines: structures within the PDRs. Red dash: Filament. Green dash: Diffuse}
	\label{fig:seds_comp}
\end{figure}

While the measurements are usually straightforward on the MIPS 24 $\mu$m profiles, they are significantly more uncertain on the MIPS 70 $\mu$m profiles, especially for less contrasted structures like the ``Filament'' or the ``Diffuse'' emission. In those two last cases, we are not sure about the exact spatial extent of the structure at 70 $\mu$m and the range over which to estimate the background (see Fig.~\ref{fig:cut_plot_diff}). This generally also applies to the IRAC 8 and 6 $\mu$m measurements, but to a lesser extent. In particular, for the ``Filament'' structure, the discrepancy in the profile's peak position between MIPS 24 $\mu$m and MIPS 70 $\mu$m or IRAC 8 $\mu$m is significant enough so we do not consider them as probing the same physical conditions (see Fig.~\ref{fig:cut_plot_fila} and \ref{fig:cut_plot_diff}). Since there is no other obvious feature at the position of the MIPS 24 $\mu$m peak, we will thus use the MIPS70 $\mu$m measurement as an upper limit. Additionally, the uncertainty on the MIPS 70 $\mu$m flux of the ``Diffuse'' is significantly higher. The resulting photometric SEDs, normalized to MIPS 24 $\mu$m, are presented on figure \ref{fig:seds_comp}. Again, the differences between the structures within the shell and those within the PDRs are clear.

\begin{itemize}
\item The PDRs of M16, both at the tip of the Spire and within the Pillars of Creation, are characterized by an almost flat SED from near to mid infrared and a continuous increase mid to far infrared wavelengths. The SEDs of the position with an embedded source (``Pillar'' and ``Spire'') do not appear to be different from that of the ``Shoulder'' at near infrared wavelengths. At longer wavelengths, the SED of the ``Shoulder'' increases slightly less than those of the ``Pillar'' and the ``Spire'', which encompass embedded source. The ratio between MIPS24 and MIPS70 is about 0.1 for the ``Shoulder'' and about 0.3 at the tip of the main Pillar of Creation and the Spire.
\item The inside shell, at Pilbratt's blob position and on bright contrasted structures, is characterized by a significantly steeper increase of the intensity from near to mid infrared and a flat or decreasing SED from mid to far infrared. On Pilbratt's blob, the Shell border and the Reverse shell, the MIPS24 to MIPS 70 ratio is about 4.5, 2.3 and 0.95 respectively.
\item The Filament and the Diffuse SEDs appear in between these two sets of SEDs. Both their MIPS24 to MIPS 70 ratio is lower than inside the shell and their near to mid infrared SED is steeper than within PDRs but the uncertainties are significantly larger. As a consequence, in the following sections, we do not discuss these last two positions. 
\end{itemize}

The measurements of the near-IR to far-IR SEDs confirm what spectroscopic observations were suggesting: the dust within the inner shell is significantly different from that within PDRs. The addition of the MIPS 70 $\mu$m and its comparison to MIPS 24 $\mu$m provide us with constraints on the position of the dust emission peak in the FIR. We explore in the next section whether the difference arises from external excitation or intrinsic properties.

\section{UV heating of the dust}
\label{lab:uvheating}

In this section, we model the dust emission within M16 using the dust model of \citet{Compiegne2011}. In this model, the dust is heated by the incident flux of UV photons only. We first show that the MIPS 24 $\mu$m to MIPS 70 $\mu$m ratio may be directly related to the intensity of the interstellar radiation field (ISRF) in the shell. We then use the dust model to determine the best set of parameters that describes the complete observed SEDs over the entire nebula. In this section, we limit ourselves to the following parameters : the intensity of the incident radiation field and the dust size distribution, in terms of abundance of the dust components.

\subsection{Method}
\label{lab:modmeth}

The dust model of \citet{Compiegne2011} is an updated version of the original \citet{Desert1990} model. In their model, \citet{Compiegne2011} use four dust components: (1) polycyclic aromatic hydrocarbons (PAH), (2) stochastically heated very small grains of amorphous carbon (VSG or SamC), (3) large amorphous carbon grains (LamC) and (4) amorphous silicates (aSil). We combine LamC and aSil grains into a unique big grains (BG) component using these grains relative abundances found in the diffuse high galactic latitude (DHGL) medium \citep{Compiegne2011}. We assume a fix dust-to-gas mass ratio of 1\%. We then use the dust model to compute the emission spectra of the three dust components (PAHs, VSGs and BGs) illuminated by the incident radiation field from the star cluster NGC6611.

We use the STARBURST99 online model\footnote{http://www.stsci.edu/science/starburst99/} described in \citet{Leitherer1999} and \citet{Vazquez2005} to define the spectral shape of the radiation field from the illuminating star cluster NGC6611. We use the following parameters: 2 millions years old cluster, Salpeter initial mass function ($dn/dM \propto M^{-2.35}$), stellar masses from 1 $M_\odot$ to 100 $M_\odot$. The modeled radiation field corresponds to $1.6\times10^9\ L_\odot$. We normalize it so that it is in agreement with the total flux of the most massive stars of the cluster.
\citet{Dufton2006} have presented an analysis of VLT-FLAMES spectroscopy for NGC6611. Their online catalogue \citep{Dufton2006cat} lists stars classified as earlier than B9. The 42 members of NGC6611 have a combined total luminosity of $3.4\times 10^6\ L_\odot$, which is a factor 480 smaller than the Starburst99 model output spectrum. We apply that correction factor to the model spectrum of the ISRF. In Habing units -- integrated intensity of the solar neighborhood from 912 to 2000 \AA\ or $1.6\times10^{-3}\ \rm{erg.s^{-1}.cm^{-2}}$ -- the cluster radiation field intensity is $\chi_0 \simeq 4800$ at a distance of 3 parsecs (see section \ref{lab:chi} for a discussion on the spatial variations of the IRSF). In the following, we use this value as a reference for the dust model.

For the features within the shell (``Blob'', ``Shell border'' and ``Reverse shell''), the use of a non-attenuated radiation field is acceptable since the UV optical depth is low. For the features within the PDRs (``Pillar'', ``Spire'' and ``Shoulder''), we have to take into account the extinction of the ISRF by the ionized layer of gas and the PDR layer itself. We model this in a simple way by removing the Lyman continuum photons and with a far-UV extinction of 1 magnitude. Such an extinction accounts for the fact that the emission from PDRs comes from a range of depths into UV-dark clouds with a weighting proportional to the UV field. A more detailed study of the PDRs is beyond the scope of this paper.

\subsection{MIPS 24 $\mu$m to MIPS 70 $\mu$m ratio as a tracer of $\chi$}
\label{lab:m24m70_vs_chi}

\begin{figure*}[!t]
\centering
\subfigure[] 
	{\label{fig:mdl_2470_lyc}
	\includegraphics[angle=90,width=.475\linewidth]{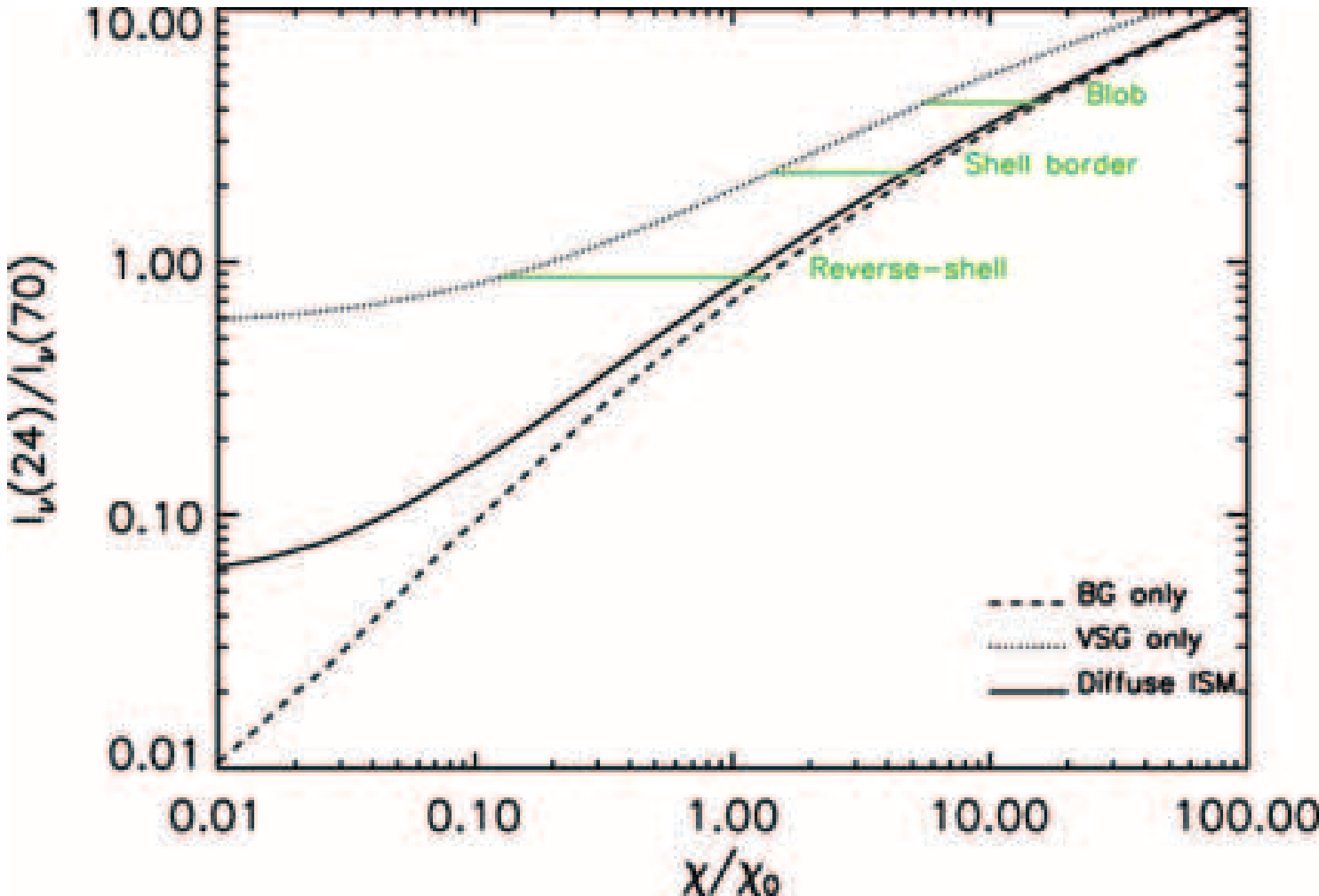}}
\subfigure[] 
	{\label{fig:mdl_2470_nolyc}
	\includegraphics[angle=90,width=.475\linewidth]{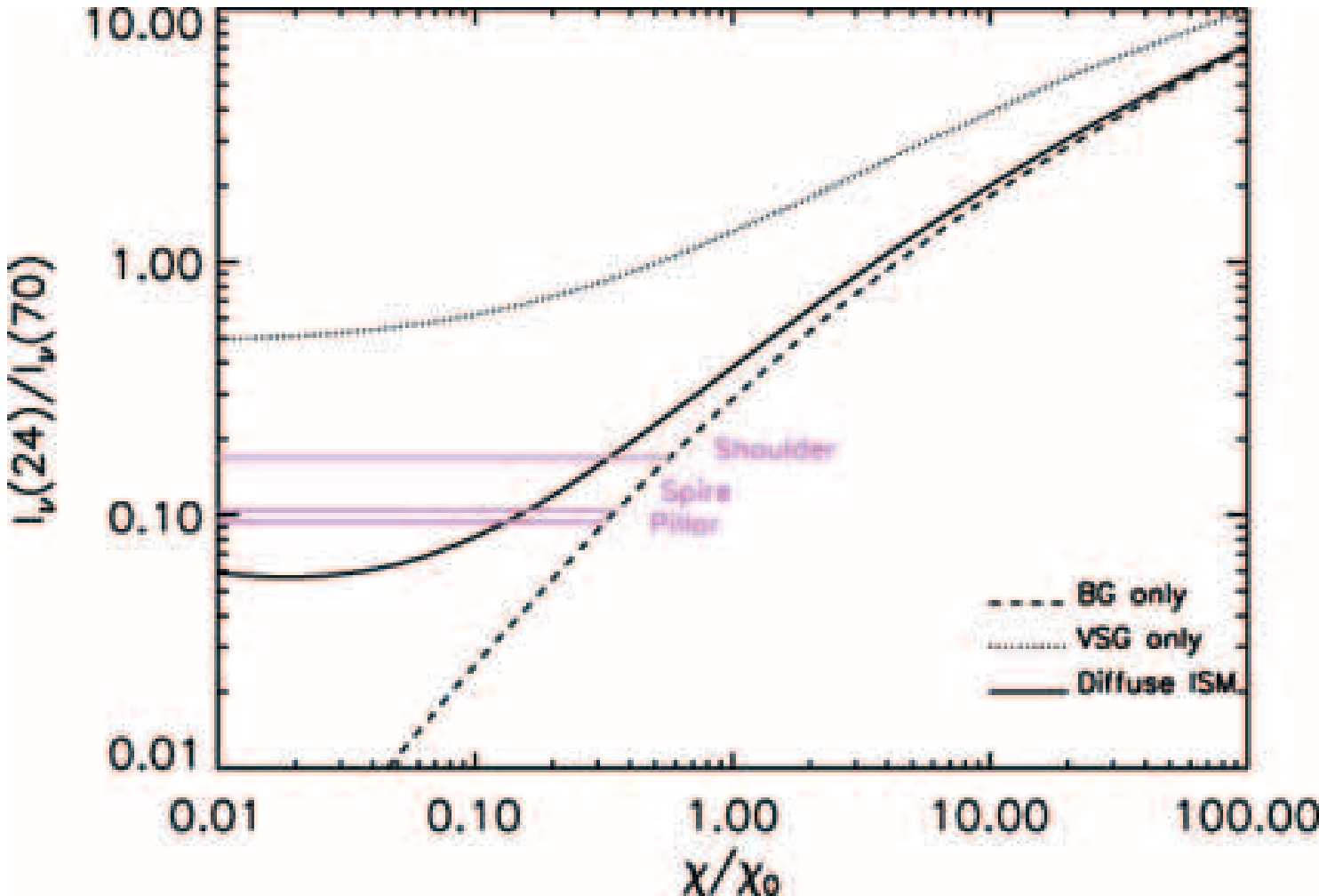}}
\caption{MIPS 24 $\mu$m to MIPS 70 $\mu$m ratio as a function of the ISRF intensity, as predicted by the model of \citet{Compiegne2011}. Several dust size distribution are used: (dotted line) BGs only, (dashed-line) VSGs only and (solid line) mix of BGs and VSGs. The MIPS24-to-MIPS70 ratio for several structures within M16 is indicated. The ISRF spectral shape is that mention in the text with (a) no extinction, (b) A(FUV) = 1 mag and the Lyman continuum photons removed.}
\label{fig:mdl_2470_nolya}
\end{figure*}

\begin{figure*}[!t]
\centering
\subfigure[] 
	{\label{fig:mdl_bg_teq_lyc}
	\includegraphics[angle=90,width=.475\linewidth]{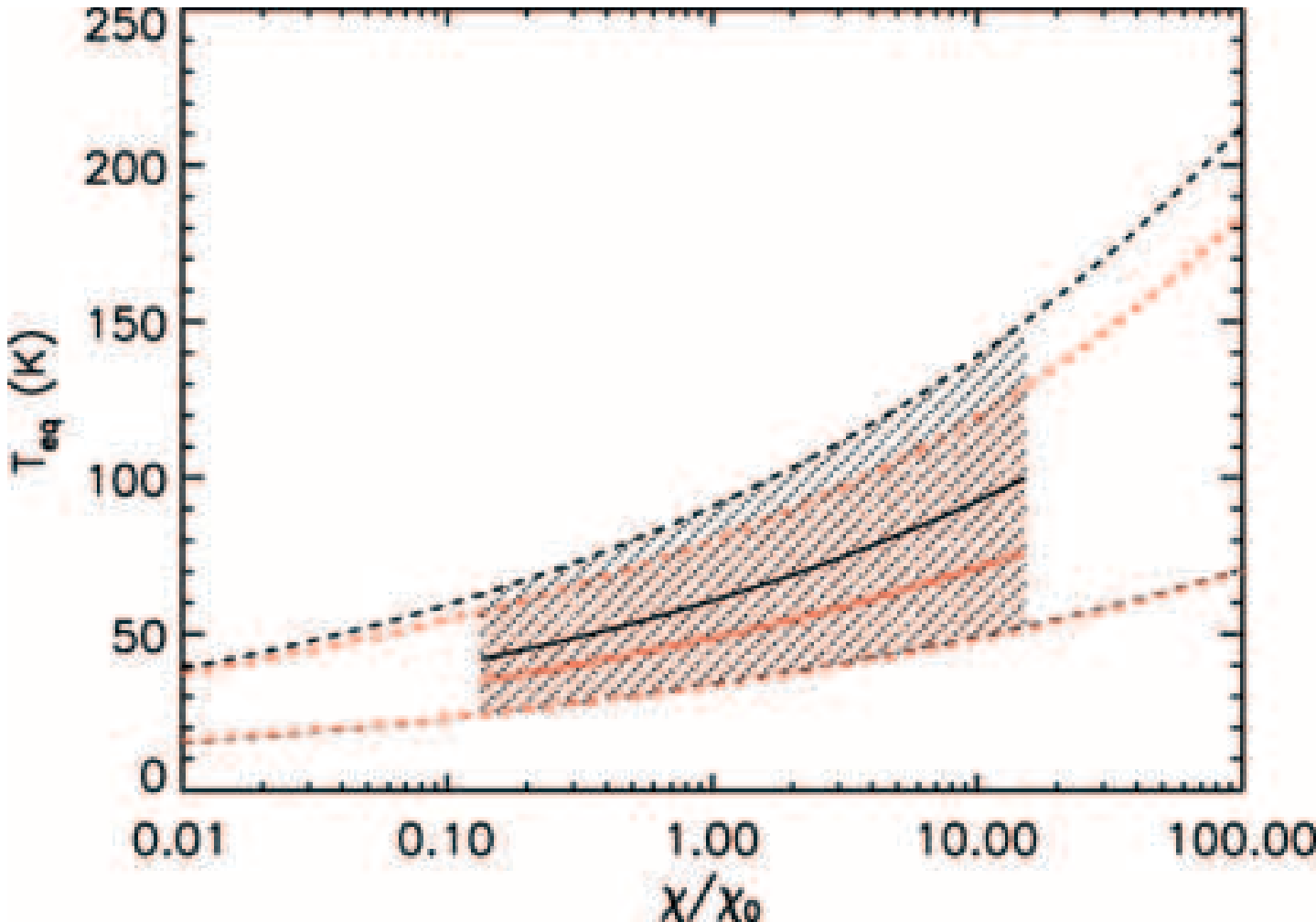}}
\subfigure[] 
	{\label{fig:mdl_bg_teq_nolyc}
	\includegraphics[angle=90,width=.475\linewidth]{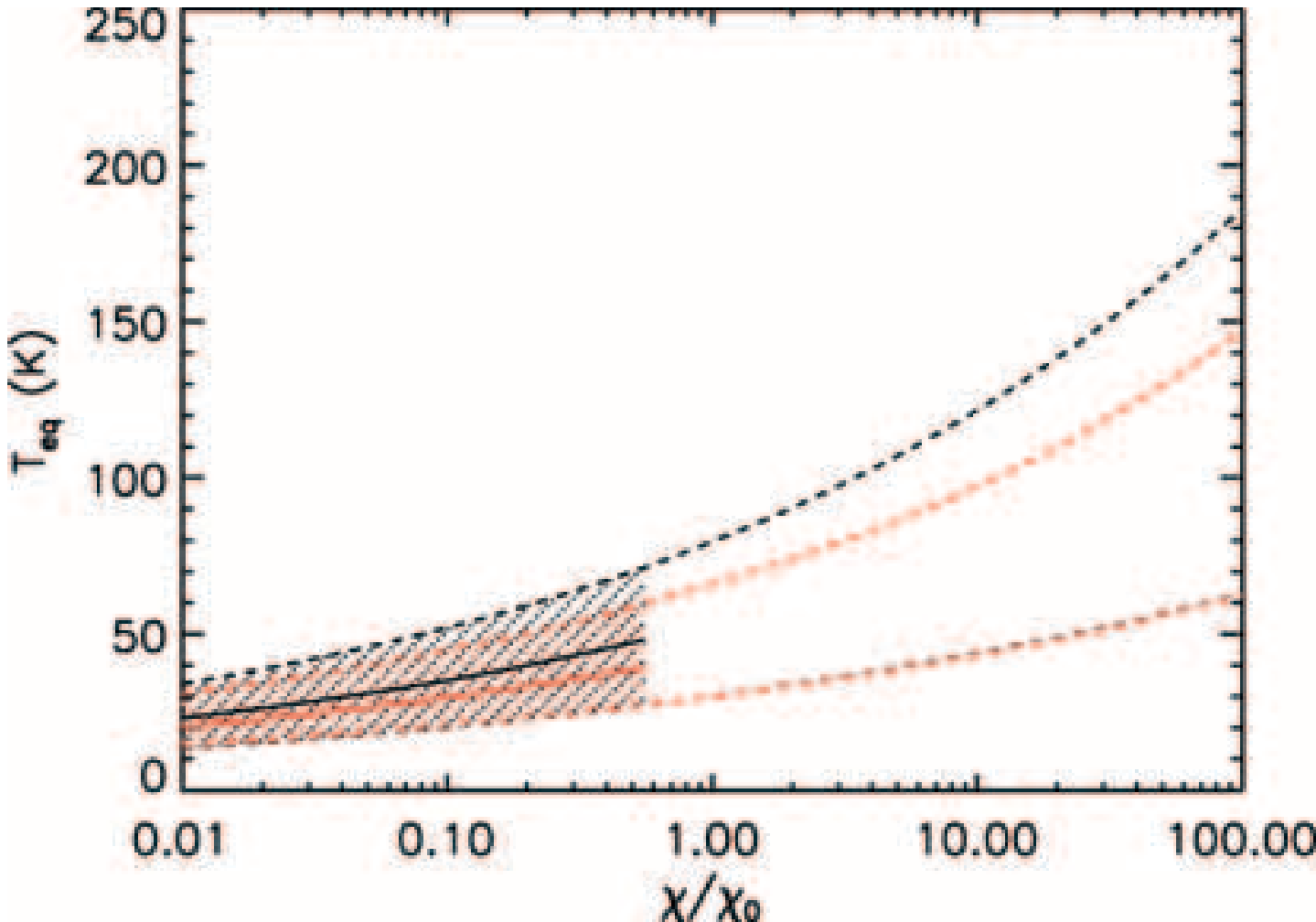}}
\caption{BGs equilibrium temperature as a function of the ISRF intensity. The hatched area corresponds to the range of equilibrium temperatures span by the entire BGs size distribution. The solid lines represent the equilibrium temperature for the most abundant size bin. The hatched area and the solid line are only plotted for the values of $\chi/\chi_0$ that are given by figure \ref{fig:mdl_2470_nolya}. Black is for LamC grains, red is for aSil grains as described in \citet{Compiegne2011}. The ISRF spectral shape is that mention in the text with (a) no extinction, (b) A(FUV) = 1 mag and the Lyman continuum photons removed.}
\label{fig:mdl_bg_teq}
\end{figure*}

We first use the dust model of \citet{Compiegne2011} to compute the MIPS 24 $\mu$m to MIPS 70 $\mu$m ratio of the dust emission for different dust size distributions to show how it is related to $\chi$. Within this wavelength range, the PAHs contribution to the emission is weak relative to that of VSGs and BGs. Therefore, we present the MIPS 24 $\mu$m to MIPS 70 $\mu$m ratio as a function of $\chi$ for three size distributions: VSGs only, BGs only and a mixture of VSGs and BGs that matches their relative abundance in the diffuse high Galactic latitude medium \citep[DHGL,][]{Compiegne2011}. Therefore, we take into account any dust evolutionary process that would destroy a specific grain size component. Figure \ref{fig:mdl_2470_nolya} shows the results along with the MIPS 24 $\mu$m to MIPS 70 $\mu$m ratio measured for the Eagle Nebula structures, both within the shell and the PDRs. The differences between the set of curves for the PDRs and that for the shell are not significant. We first make no distinction while presenting them. Then we discuss the results for the PDRs and Shell structures independently.

For a given $\chi/\chi_0$, VSGs always have a higher MIPS24/MIPS70 as they are hotter than BGs. However, for $\chi/\chi_0 \gtrsim 1.0$, MIPS24/MIPS70 is almost independent, within a factor of a few, from the dust size distribution. These values of $\chi$ correspond to the large values of the MIPS24/MIPS70 ($>1$). For $\chi/\chi_0 \lesssim 1.0$, MIPS24/MIPS70 is significantly more dependent on the grain size distribution with difference up to almost 2 orders of magnitude. Likewise, for a given MIPS24/MIPS70, the required $\chi/\chi_0$ is always higher for BGs than VSGs. The difference is as small as a factor of a few for high values of MIPS24/MIPS70 and as high as almost 2 orders of magnitude for low values of MIPS24/MIPS70. Therefore, given MIPS24/MIPS70, the constraint on the intensity of the IRSF is stronger for higher values of $\chi$ and requires a better knowledge of the dust size distribution (e.g. as provided by other IR observations, see next subsection) at low values of $\chi$. On the contrary, constraining the dust size distribution requires an \textit{a priori} on $\chi$ and can better be done at low values of $\chi$.

According to the model, the PDR structures (``Pillar'', ``Spire'' and ``Shoulder'') require an ISRF intensity at most a factor 2 lower than the reference, and no lower limit can be estimated because we have no constraint on the dust size distribution. However, if we assume it does not significantly depart from that of the DHGL medium, the MIPS24 to MIPS70 ratio within PDRs are best interpreted with $\chi/\chi_{ISRF,0} \simeq 0.1$. The inner shell structures (``Blob'', ``Shell border'' and ``Reverse shell'') are more on the high end of the ISRF intensity. The ``Shell border'' and the ``Blob'' are best interpreted with a $\chi/\chi_{ISRF,0}$ of at least a few and up to 16, whether the dust size distribution is dominated by BGs or VSGs. The difference between the ISRF intensity that illuminates these two structures and the PDRs is thus at least an order of magnitude. The ``Reverse shell''' position however is not strongly constrained and overlaps those of the PDRs structures. If at this position the dust size distribution is dominated by VSGs, then $\chi /\chi_{ISRF,0} \simeq 0.1$ while $\chi/\chi_{ISRF,0} \sim 1$ if the BGs contribute the most to the dust size distribution. The full range of required ISRF intensities for each structure is given in Table \ref{tab:minmaxchi}.

\begin{table}[!hb]
\caption{Lower and upper limits of $\chi/\chi_0$ for the whole set of structures as deduced from their MIPS 24 $\mu$m to MIPS 70 $\mu$m ratio.}
\begin{center}
\begin{tabular}{l l  |  l l}
\hline
\hline
Shell structure & $\chi/\chi_0$ & PDR structure & $\chi/\chi_0$ \\
\hline
Reverse shell & 0.13-1.3 & Pillar & $< 0.3$ \\
Blob & 5.6-16 & Shoulder & $< 0.6$ \\
Shell border & 1.4-5.4 & Spire & $< 0.4$ \\
\hline
\end{tabular}
\end{center}
\label{tab:minmaxchi}
\end{table}

Indirectly the MIPS 24 $\mu$m to MIPS 70 $\mu$m ratio also provides us with a measurement of the equilibrium dust temperature $T_{eq}$ of the largest dust particles. In figure \ref{fig:mdl_bg_teq}, we plot the BGs equilibrium temperature, provided by the dust model, as a function of $\chi$, for both the PDR and the Shell structures, and for both types of large grains used in the model of \citet{Compiegne2011}: LamC and aSil. For a given radiation field intensity $\chi/\chi_{ISRF,0}$, we plot the upper and lower limits for the equilibrium temperatures of each grain type. The difference between both type of BG components is not really significant. In figure \ref{fig:mdl_bg_teq}, we hatch the range of equilibrium temperatures for the values of $\chi/\chi_0$ given by Fig.~\ref{fig:mdl_2470_nolya}: $0.13 <\chi/\chi_0< 16$ for the Shell and $\chi/\chi_0< 0.6$ for the PDR structures. While the smallest LamC grains in the PDR structures may reach equilibrium temperature as high as 71 K, those are in limited number. Likewise, only the largest grains in the Shell structures may reach equilibrium temperature as low as 24 K. The majority of the grains, as traced by the most abundant size bin of each BG component (also plotted in figure \ref{fig:mdl_bg_teq}), span a range of equilibrium temperatures that does not overlap significantly between the Shell and the PDR structures. For the PDRs structures, equilibrium temperatures for the most abundant size $20 < T_{eq} < 50 K$ while for the inner shell structures $35 K < T_{eq} < 100 K$. Therefore, equilibrium temperatures above 50 K can only be efficiently reached by BGs in the inner shell while equilibrium temperatures below 50 K are mostly found in the PDRs. The dust in the inner shell is thus significantly hotter than that in the PDRs. \citet{Indebetouw2007} have used IRAS 60 $\mu$m to IRAS 100 $\mu$m ratio to build a low spatial resolution (4.3\arcmin) color temperature map of the dust in M16. Their values range from 32 K in the molecular cloud to 40 K inside the nebula. We build the same map (not shown here) using IRAS 25 $\mu$m to IRAS 60 $\mu$m ratio (to better match the MIPS24 to MIPS 70 $\mu$m diagnostic) and find color temperature ranging from 45 K to 65 K, more in agreement with our measurements of the BGs equilibrium temperature in the shell. The remaining difference may come from the lower spatial resolution that averages ``hot'' features with ``cold'' features within the beam.

\subsection{Fitting of the whole IR SED}
\label{lab:fit}

\begin{figure*}[!t]
\centering
\subfigure[Pillar]
	{\label{fig:head_sed_plot_fit}
	\includegraphics[angle=90,width=.3\linewidth]{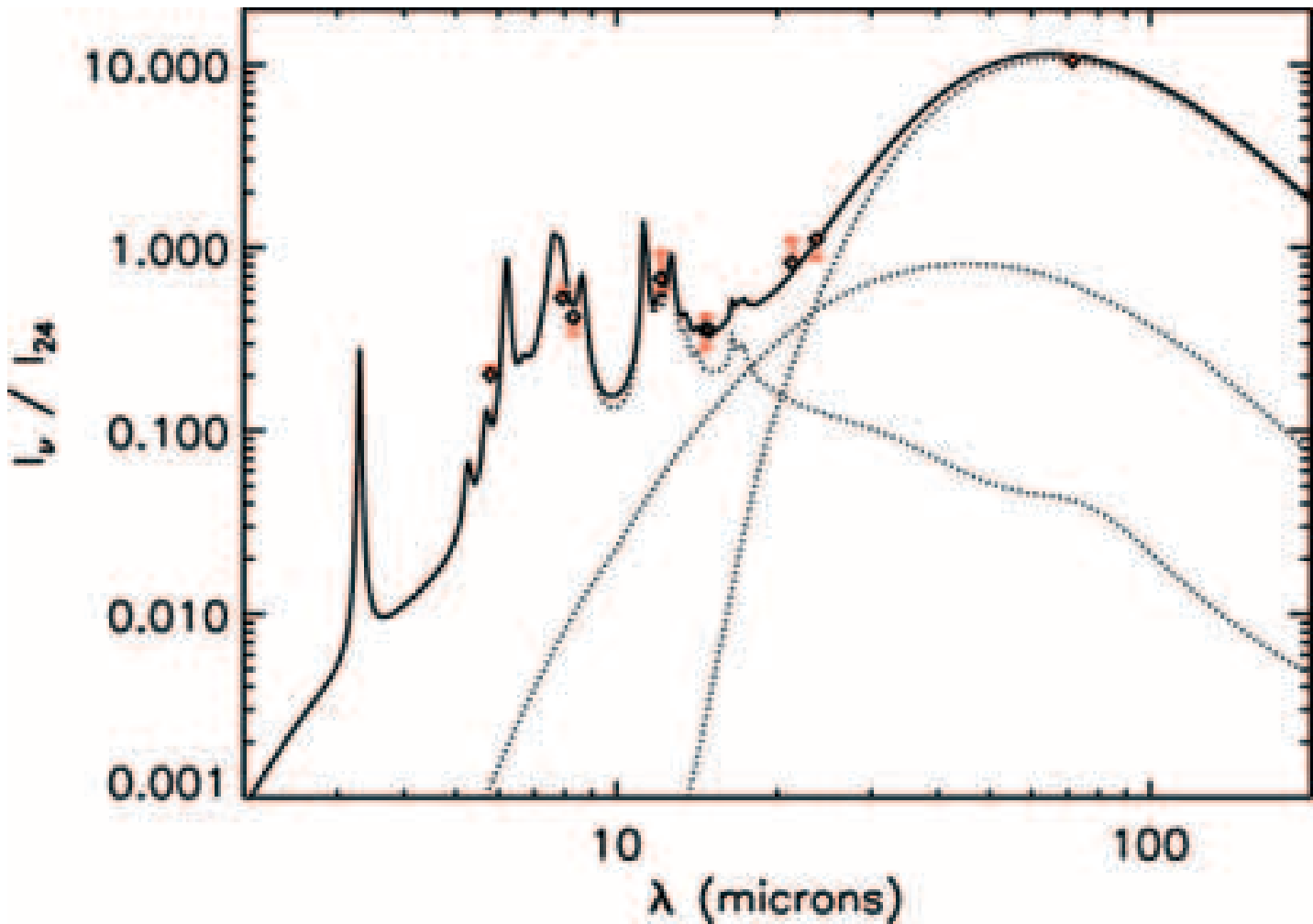}}
\subfigure[Shoulder]
	{\label{fig:pdr_sed_plot_fit}
	\includegraphics[angle=90,width=.3\linewidth]{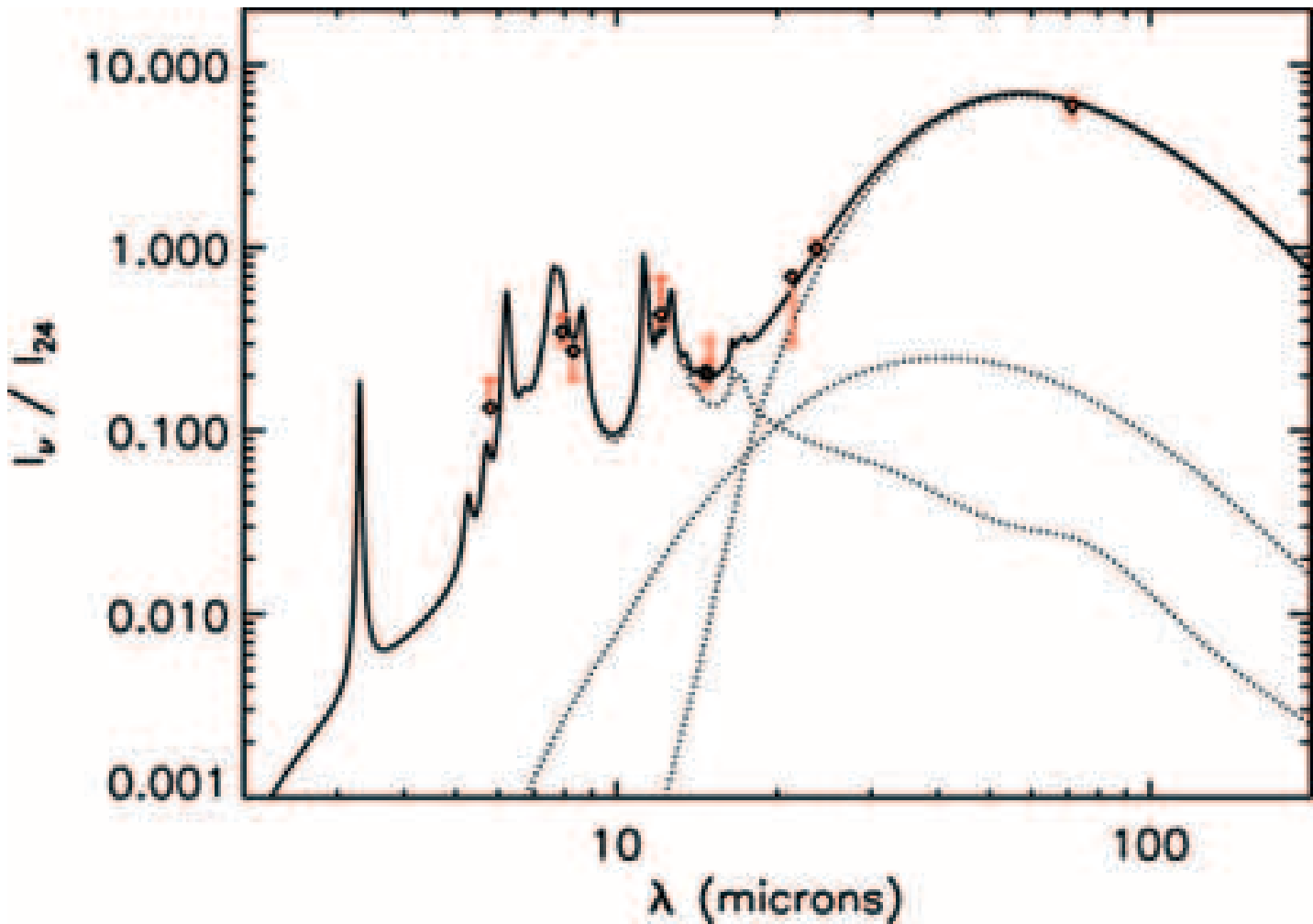}}
\subfigure[Spire]
	{\label{fig:fairy_sed_plot_fit}
	\includegraphics[angle=90,width=.3\linewidth]{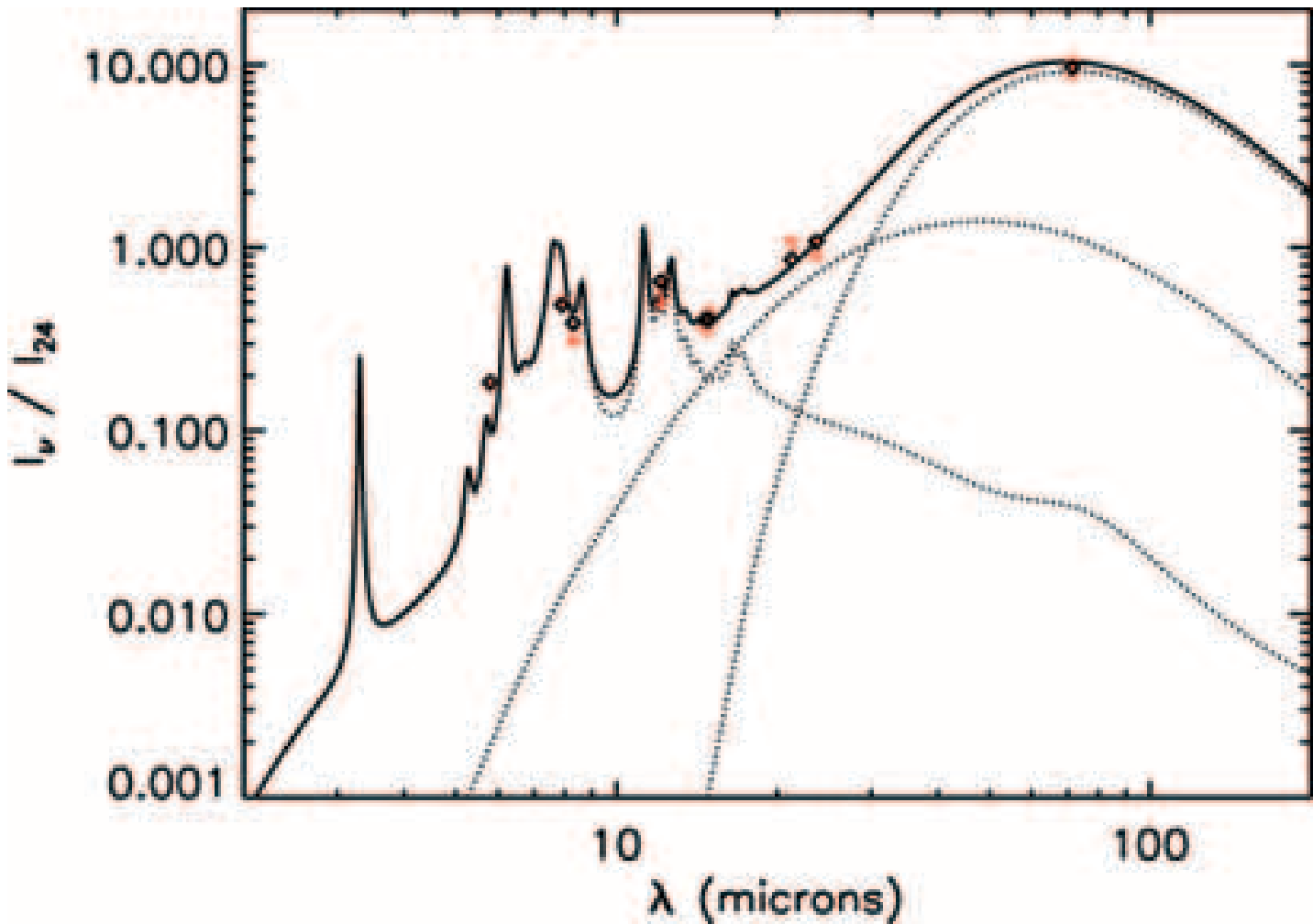}}
\subfigure[Shell border]
	{\label{fig:shell_sed_plot_fit}
	\includegraphics[angle=90,width=.3\linewidth]{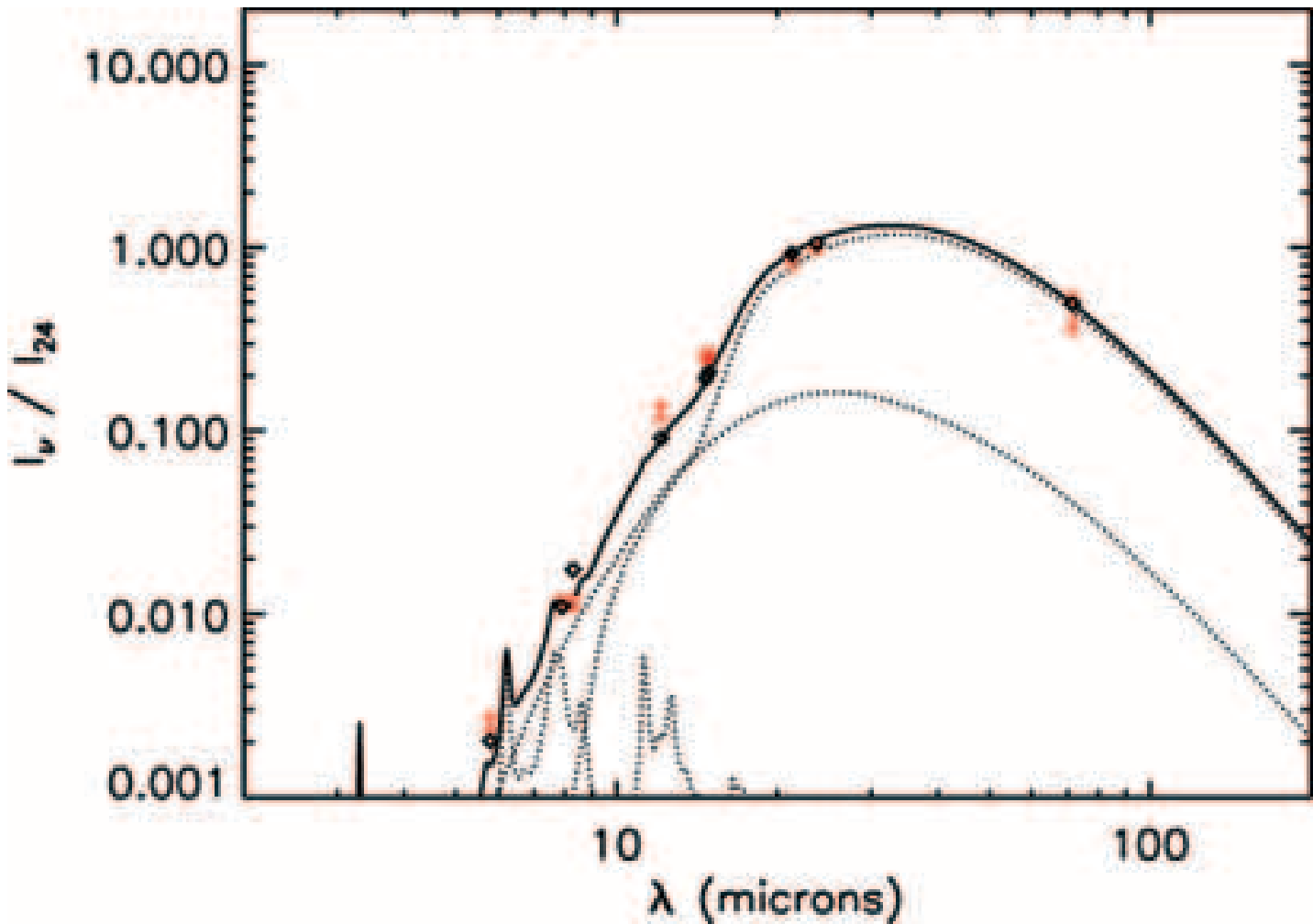}}
\subfigure[Blob]
	{\label{fig:blob_sed_plot_fit}
	\includegraphics[angle=90,width=.3\linewidth]{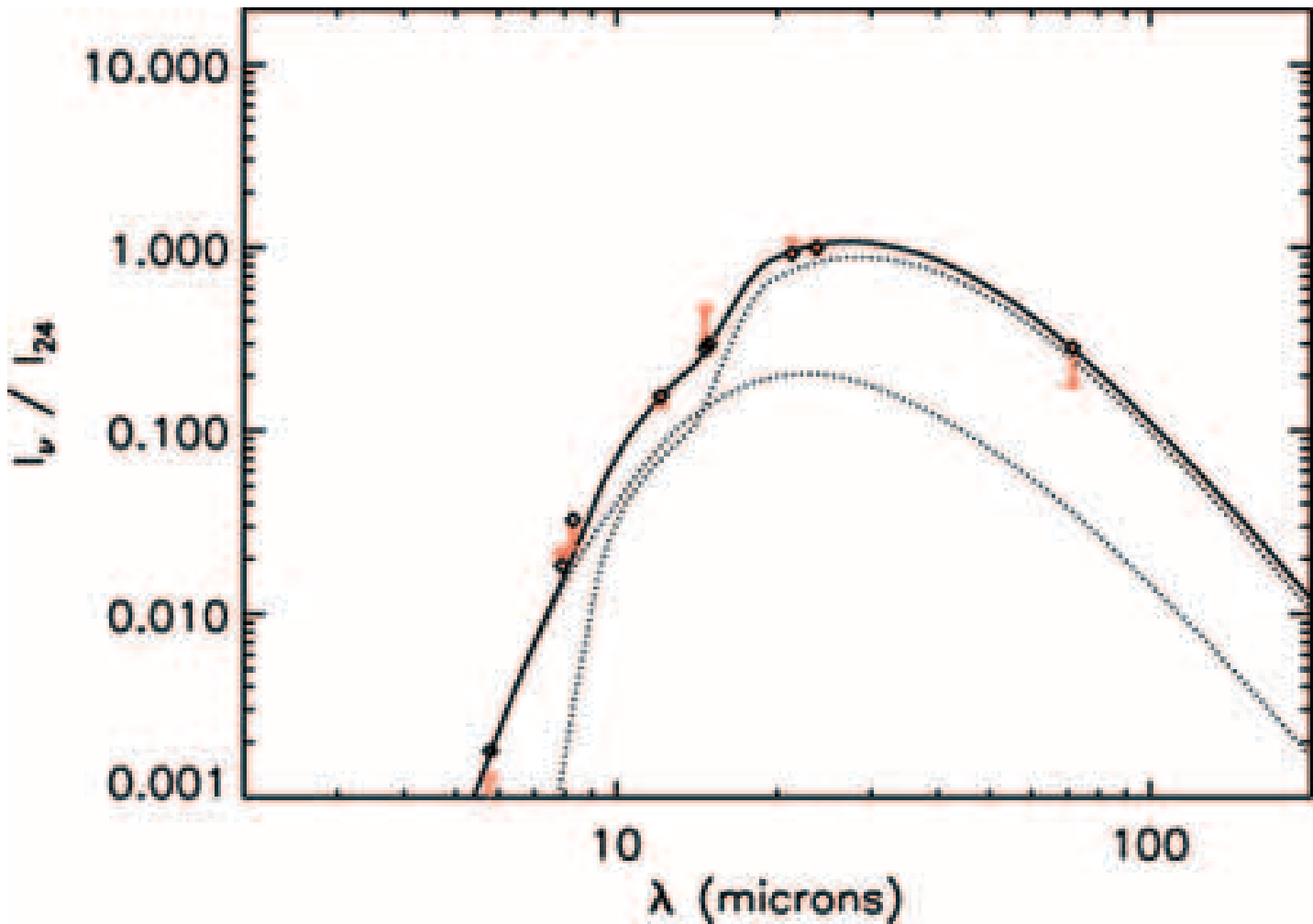}}
\subfigure[Reverse shell (no IRAC5.8)]
	{\label{fig:backshell_sed_plot_fit_b}
	\includegraphics[angle=90,width=.3\linewidth]{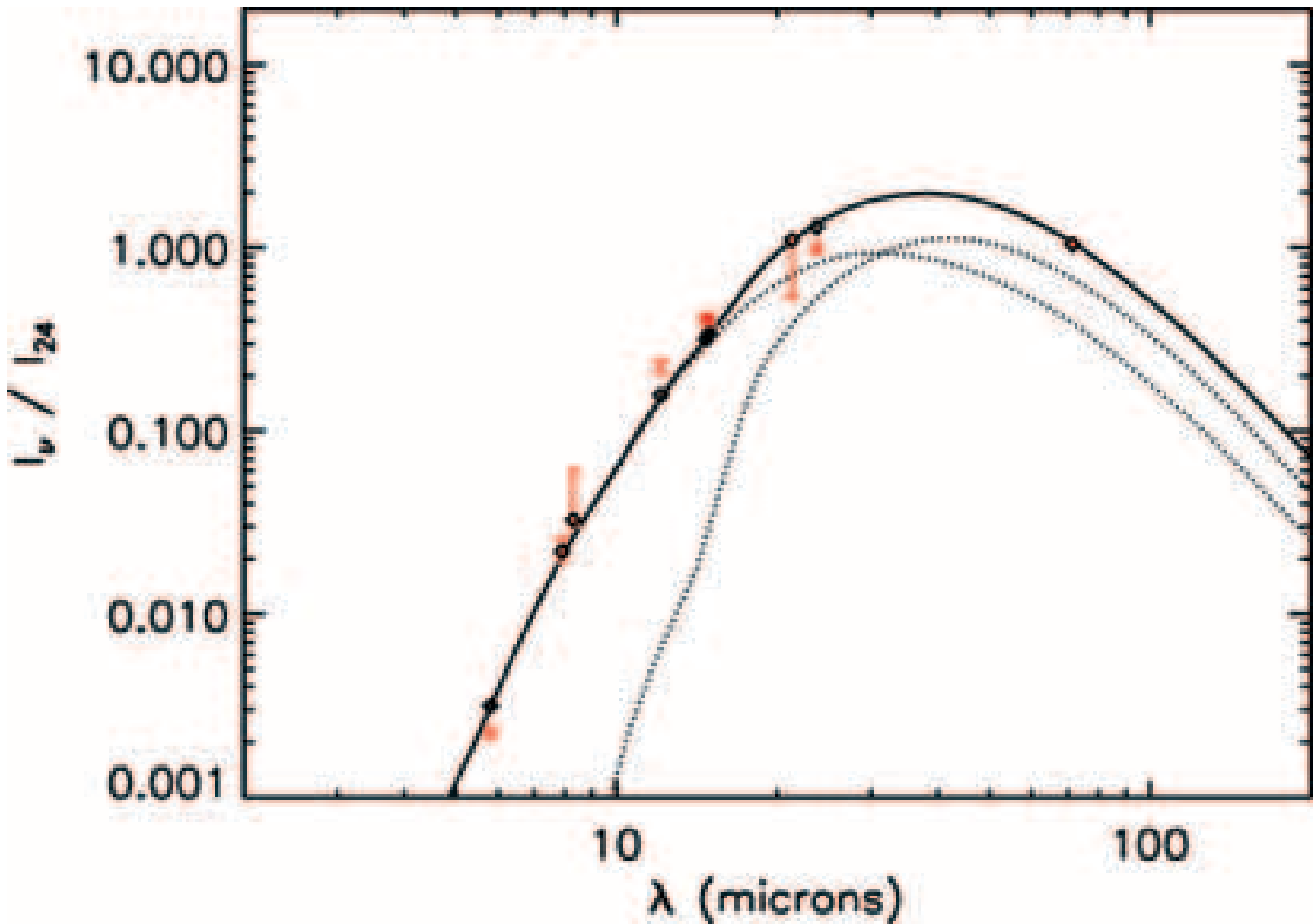}}
\caption{Best-fit for (a) the ``Pillar'', (b) the ``Shoulder'', (c) the ``Spire'', (d) the ``Shell border'', (e) the ``Blob'', (f) the ``Reverse shell''. Solid line is the total model spectrum, dotted-lines are PAHs, VSGs, and BGs contributions. Diamonds are model broadband fluxes. Red crosses are measurements.}
\label{fig:sed_plots_fits}
\end{figure*}

\begin{table*}[!]
\caption{Best-fit parameters for SEDs of the Eagle Nebula. The ISRF intensity, the dust size distribution, in terms of relative mass ratio abundances, and the total dust column density are given. The parameters for the diffuse high Galactic latitude (DHGL) reference of \citet{Compiegne2011} are also given. The dust-to-gas mass ratio is fixed at 0.01 therefore a dust mass column density of $1.7 \mu g.cm^{-2}$ corresponds to $10^{20}\ \rm{H.cm^{-2}}$.}
\begin{center}
\begin{tabular}{l  c  c  c  c  c}
\hline
\hline
Position & $\chi/\chi_0$ & $Y_{PAH} (M/M_H)$ & $Y_{VSG} (M/M_H)$ & $Y_{BG} (M/M_H)$ & $\sigma_{dust}\ \rm{(\mu g.cm^{-2})}$ \\
\hline
DHGL              &                               & $7.8 \times 10^{-4}$ & $1.65 \times 10^{-4}$ & $9.25 \times 10^{-3}$ & 1.7 \\
\hline
Pillar        &  $0.19 \pm 0.04$  &  $ (2.64 \pm 0.57)\times10^{-4}$  &  $ (2.45 \pm 0.90)\times10^{-4}$  &  $ (9.49 \pm 1.82)\times10^{-3}$  &  380 \\
Shoulder  &  $0.43 \pm 0.08$  &  $ (2.51 \pm 0.45)\times10^{-4}$  &  $ (1.12 \pm 0.95)\times10^{-4}$  &  $ (9.64 \pm 2.15)\times10^{-3}$  &   33 \\
Spire       &  $0.12 \pm 0.05$  &  $ (2.96 \pm 1.27)\times10^{-4}$  &  $ (5.09 \pm 2.89)\times10^{-4}$  &  $ (9.20 \pm 3.62)\times10^{-3}$  &  870 \\
\hline
Shell border    &  $4.36 \pm 1.36$  &  $ (4.85 \pm 1.12)\times10^{-6}$  &  $ (3.69 \pm 2.71)\times10^{-4}$  &  $ (9.63 \pm 2.77)\times10^{-3}$  &    0.2 \\
Blob                &  $9.69 \pm 2.33$  &  $0$  &  $ (5.98 \pm 3.07)\times10^{-4}$  &  $ (9.40 \pm 1.82)\times10^{-3}$  &    2.9 \\
Reverse shell  &  $1.15 \pm 0.13$  &  $0$  &  $ (1.99 \pm 0.31)\times10^{-3}$  &  $ (8.01 \pm 0.23)\times10^{-3}$  &    2.1 \\
\hline
Shell Border & 2* & $(6.68 \pm 4.47)\times 10^{-5}$ & $(1.05 \pm 0.77) \times 10^{-2}$ & $(3.59\pm2.29)\times 10^{-4}$ & 0.17 \\
$a_0(VSG) = 5.5$ nm \\
\hline
\end{tabular}
\end{center}
\label{tab:sed_fit_param}
\end{table*}

The additional measurement provided by MIPS 70 $\mu$m enables us to give some constraint on the ISRF intensity that is required to heat the dust up to the observed temperatures. Hereafter we use our dust model and the whole IR SED of each structure within M16 to better determine the variation of $\chi$ and the dust size distribution at the same time.

We set five parameters free: the intensity of the ISRF and the abundances of the three dust components (PAH, VSG and BG). The spectral shape of the ISRF is that described in section \ref{lab:modmeth}. The other parameters describing the dust size distribution (e.g. the size range and distribution shape) are those presented in \citet{Compiegne2011}.
We use the MPFIT package\footnote{http://purl.com/net/mpfit} for IDL \citep{Markwardt2009} to constrain the free parameters, given the SED. We use the default tolerance parameters and limit the four parameters to positive values.
The best-fit spectra are shown on Fig.~\ref{fig:sed_plots_fits} and the best-fit parameters are given in table \ref{tab:sed_fit_param}. We also give an estimate of the dust column density for each feature \textbf{assuming a dust-to-gas mass ratio of 0.01}.

The positions within the PDRs are best fit with low values of $\chi/\chi_0$ (a few $10^{-1}$) in agreement with the upper-limits from Table \ref{tab:minmaxchi}, a factor of a few less PAHs and a factor of a few more or less VSGs than in the DHGL medium. An increase/decrease of the small grains abundance by a factor of a few within PDRs of NGC2023N and the Horsehead Nebula has already been observed by \citet{Compiegne2008} and within translucent sections of the Taurus Molecular Complex by \citet{Flagey2009}. The low values of $\chi/\chi_0$ required to fit the SED of the PDRs can partly be explained with shadow effects within the nebula. Another parameter that we do not take into account in our simple model is the geometry of the features and the resulting limb brightening effect. Indeed, the dust column density for the ``Pillar'' and the ``Spire'' position is about a few $10^{-4}\ \rm{g.cm^{-2}}$ which corresponds to a gas column density of a few $10^{22}\ \rm{cm^{-2}}$ or a visual extinction of a few magnitudes, significantly larger than that required for attenuating the incident UV radiation field. The three ``PDR'' positions give very similar results, especially in terms of PAH abundance which varies by less than 10\%. The VSG abundance is varying more significantly, up to a factor 5. The BG always dominates the dust size distribution with abundance very close to that of the DHGL.

The positions within the ``Shell'' require larger values of $\chi/\chi_0$ (about a few) in agreement with values from Table \ref{tab:minmaxchi}, a significant depletion of the PAHs and a significant increase of the VSG abundance, up to a factor 10, with respect to the PDRs values and at the expense of the BG component. The total dust column density is about $\sim 10^{-6}\ \rm{g/cm^2}$, simliar to DHGL values and which corresponds to a gas column density of about $10^{20}\ \rm{cm^{-2}}$. As a consequence of the increased $\chi$, the VSG and BG emission spectra peak at very close wavelengths (see Fig.~\ref{fig:shell_sed_plot_fit}, \ref{fig:blob_sed_plot_fit}, and \ref{fig:backshell_sed_plot_fit_b}). We show in the previous section that MIPS24/MIPS70 is a good tracer of $\chi$ but not of the dust size distribution, especially at high values of $\chi$. Here, the addition of the other IR observations provides better constraints on the VSGs to BGs relative abundance. For the position of the ``Reverse shell'', the initial best-fit (not shown here) underestimates the MIPS 70 $\mu$m measurement by almost an order of magnitude. As a consequence, the required $\chi/\chi_0$ is overestimated relative to that from Table \ref{tab:minmaxchi} derived from the MIPS24 to MIPS70 ratio. We believe this poor fit at the longer wavelengths is due to the uneven number of measurement at short and long wavelengths, relative to the peak of the dust emission. From 6 to 24 $\mu$m, no less than seven measurements are available while only MIPS 70 $\mu$m is available at wavelengths longer than the peak position. The fit process is thus biased towards shorter wavelengths. In order to limit this effect, we repeat the fitting process of the ``Reverse Shell'' position with an increased weight on the MIPS 70 $\mu$m measurement. Figure \ref{fig:backshell_sed_plot_fit_b} shows the result of that fit. The three positions within the shell give results that are very similar to each other and very different from those of the PDRs positions: (1) an incident radiation field intensity a factor of a few larger than that provided by the star cluster NGC6611 and about an order of magnitude larger than that required for the PDR positions, (2) a significant depletion of the PAHs and (3) an increase of the VSGs abundance relative to BGs as compared to the PDRs positions.

\begin{figure}
\centering
\includegraphics[angle=90,width=\linewidth]{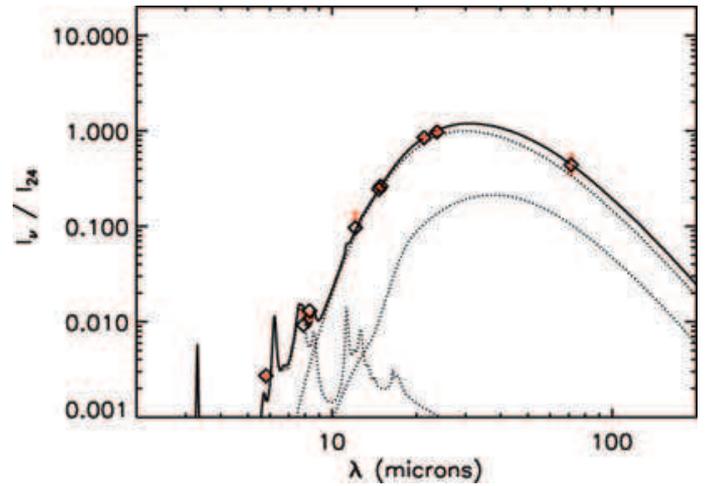}
\caption{Same as Figure \ref{fig:shell_sed_plot_fit} but with $\chi/\chi_0 = 2$ and a free mean size of the VSG component.}
\label{fig:otherfit}
\end{figure}

\textbf{In order to explore furthermore the importance of a change in the dust size distribution, we redo the fit of the ``Shell border'' with a fixed intensity of the radiation field $\chi/\chi_0 = 2$ and a free mean size of the VSG component ($a_0$). In the model of \citet{Compiegne2011} for the DHGL medium, the VSGs size distribution is assumed to have a log-normal distribution (with the centre radius $a_0 = 2$ nm and the width of the distribution $\sigma = 0.35$ nm). We keep the width of the log-normal distribution constant and set free the centre size $a_0$ between 0.6 and 20 nm. The other free parameters for that fit are the abundances of the dust components, as previously. The best-fit is plotted in Figure \ref{fig:otherfit} and the parameters are given in Table \ref{tab:sed_fit_param}. A significant increase of the mean size of the VSGs, by almost a factor 3, is required. There are almost no PAHs, as in the previous fits. The BGs are about a factor 3 less abundant than in the previous fit and about a factor 30 less abundant than in the DHGL. The abundance of VSGs is about 60 times higher than in the DHGL medium, though the uncertainty remains large ($\sim 75\%$). Therefore, the ``Shell border'' SED requires that most of the dust mass is concentrated into the VSGs component. Despite those variations of the dust size distribution, the total dust column density remains very similar to that of the fit with a fixed mean size for VSGs (0.17 instead of 0.20 $\rm{\mu g.cm^{-2}}$). We also try the same fit with $\chi/\chi_0 = 1$ but find that the uncertainties on the parameters are then significantly higher ($> 100\%$).}

\textbf{We conclude that the MIR shell SED can either be accounted for a significant change in the dust size distribution or by an additional source of heating besides the star cluster radiation field. In the following, we first discuss two sources of UV heating that may account for the values of $\chi/\chi_0 > 1$ required to fit the ``Shell'' SEDs. The first one is related to the spatial variations of $\chi$ due to the exact positions of the OB stars in the sky. The second originates in the Lyman $\alpha$ photons emitted by the hydrogen and absorbed by the dust grains. We then consider, in the next section, another heating process originating from collisions with the gas.}


\subsection{Spatial variations of the incident radiation field}
\label{lab:chi}

Depending on the exact positions of the main OB stars of NGC6611 within the Eagle Nebula, the local incident radiation field intensity may vary and thus explain the required values of $\chi/\chi_0$. For the``cold'' PDRs features, it is easy to explain values of $\chi / \chi_0 < 1$ as the stars are not all together on the plane of the sky, additionally to probable shadow effects already mentioned. However, the required values of $\chi/\chi_0 > 1$ for the ``Shell'' structures cannot be accounted for by the same interpretation. In figure \ref{fig:m16}, we indicate the position and the spectral type of the members of NGC6611, according to \citet{Dufton2006cat}. We compute the variations of the ISRF intensity $\chi_0$ as a function of the position, taking into account the luminosity and position of each individual member of the cluster. We assume that all the stars and the ``Shell'' structures are in the same plane of the sky. Therefore, the values of the local ISRF intensity we compute are thus upper-limits and the corrected values of $\chi/\chi_0$ required for the best-fits are lower-limits. All these values are reporte in Table \ref{tab:corr_chi}. The corrections factors are about a factor of a few at most. The required values of $\chi/\chi_0$ for the Shell Border and the Blob are still at least a factor 2 to 3 higher than that provided by the star cluster.

\begin{table}[b]
\caption{Correction factors on $\chi_0$ from the dispersion of the stars in the sky plane and corrected $\chi/\chi_0$ required for the best-fits.}
\begin{center}
\begin{tabular}{l l l}
\hline
\hline
Position		& Correction	& Corrected $\chi/\chi_0$ \\
			& factor		& (best fit) \\
\hline
Shell border	& $<1.5$		& $>2.9$ \\
Blob			& $<4.5$		& $>2.1$ \\
Reverse shell	& $<6.8$		& $>0.2$ \\
\end{tabular}
\end{center}
\label{tab:corr_chi}
\end{table}%

\begin{figure}[]
	\centering
	\includegraphics[width=.9\linewidth]{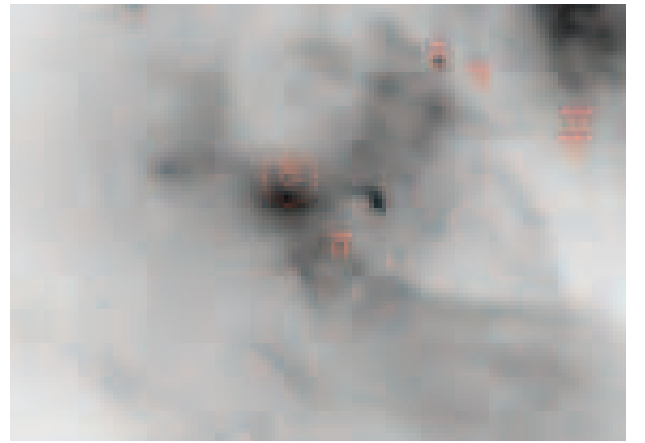}
	\caption{Blow out of the MIPS 24 $\mu$m image at the position of the Blob. The position and spectral type of O stars from NGC6611 are also reported. The red dashed circle, centered on the 08.5V star has a 26 arcsec radius (about 0.25 pc at the distance of M16).}
	\label{fig:blobowshock}
\end{figure}

The position of the members of NGC6611 also reveals that Pilbratt's Blob is very close to an 08.5V star, as shown on Fig.~\ref{fig:blobowshock}. This suggest a possible local action of the winds from this star. The shock provided by the winds may account for a local enhancement of the density within the shell and possibly for dust processing. The same interpretation does not hold for the ``Shell border'' and the ``Reverse shell'' position which both are away from any OB star, as also shown in Fig.~\ref{fig:blobowshock}. We discuss collisional heating in section \ref{lab:collheating}.

\subsection{Lyman alpha photons heating}
\label{lab:alpha}

We show here that Lyman $\alpha$ photons are not a significant heat source for the shell. Every Lyman $\alpha$ photons emitted by an hydrogen atom, after multiple absorption and reemission by other hydrogen atoms, either succeed to escape the medium or is absorbed by a dust grain. The Lyman $\alpha$ contribution to the dust IR brightness is $S_{Ly\alpha} = \int n_e \times n_{H^+} \times a_{2} \times h\nu_{Ly\alpha} dl = EM\times a_2 \times h\nu_{Ly\alpha}$, where EM is the emission measure and $a_2$ the hydrogen recombination coefficient to levels 2 and higher. The equation assumes that all recombinations from excited levels produce a Ly$\alpha$ photon that is absorbed by dust.

We compute the EM from Br$\gamma$ observations of M16 obtained at the \textit{Canada-France-Hawaii Telescope} (CFHT). These observations will be presented in a future paper. They do not show a counterpart of the ``Blob'', but there is an increase of the Br$\gamma$ emission associated with ``Shell border'' of $EM = 3.5\times10^3\ \rm{pc.cm^{-6}}$. The Ly$\alpha$ photons total flux that we estimate from these measurements is $S_{Ly\alpha} = 0.048\ \rm{erg.s^{-1}.cm^{-2}}$. In comparison, the 24 $\mu$m brightness of the ``Shell border'' is 230 MJy/sr which corresponds to a bolometric intensity of $0.37\ \rm{erg.s^{-1}.cm^{-2}}$ that we measure on the best fit (see Fig.~\ref{fig:cut_plot_shellborder}) between 1 and 1000 $\mu$m. The extra heating provided by the Ly$\alpha$ photons is thus about a factor 8 too small.

\section{Collisional heating of dust}
\label{lab:collheating}

In this section, we face the difficulty of explaining the shell infrared colors with UV heating by considering the possibility that gas-grains collisions provide additional dust heating. We quantify the conditions that would be required to fit the shell SED with a combination of radiative + collisional heating of dust.

\begin{figure}
\centering
\includegraphics[width=\linewidth]{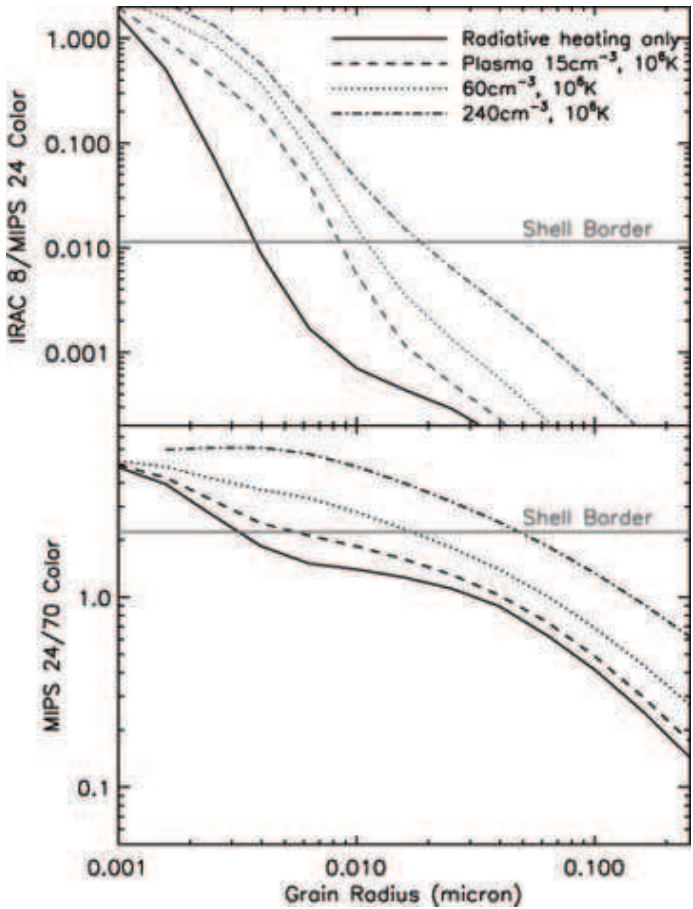}
\caption{Spitzer colors I$_\nu (8\mu$m)/I$_\nu (24\mu$m) and I$_\nu (24\mu$m)/I$_\nu (70\mu$m) for carbon grains versus grain size. The solid line give the colors for radiative heating for the Eagle Nebula ISRF. The other lines show the impact of collisional heating for a range of plasma pressure and a fixed temperature $T_e$ of $10^6$ K. A good fit of the shell SED is obtained for $n_e \times T_e \sim 3\times10^7 \, {\rm K\, cm^{-3}}$ (see Fig.~\ref{fig:shell_fit_collisions}.)}
\label{fig:IR_colors}
\end{figure}

\begin{figure}
\centering
\includegraphics[width=\linewidth]{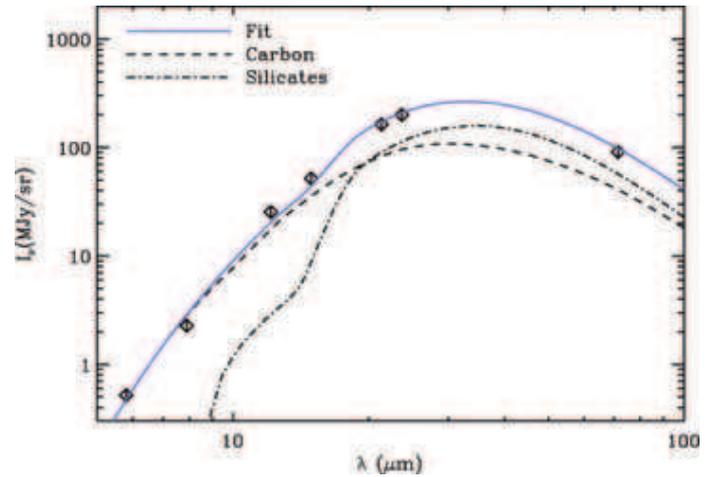}
\caption{Fit the spectral energy distribution measured on the Eagle shell with radiative plus collisional heating. The ISRF is that determined in section \ref{lab:uvheating} with $\chi/\chi_0 = 1$. The electron density is $30\ \rm{cm^{-3}}$ and the plasma temperature $10^6$ K.}
\label{fig:shell_fit_collisions}
\end{figure}

We use the work of \citet{Dwek1987} to quantify the heat deposited in the grain by collisions with electrons as a function of grain size and plasma temperature. Like in section \ref{lab:uvheating}, we use the DUSTEM model with a combination of silicates and amorphous carbon grains \citep{Compiegne2011}. Since the DUSTEM code does not include collisional excitation, we wrote a specific module to compute the distribution of grain temperatures for stochastic heating by both photons and collisions. This code takes into account the Maxwellian distribution of the electrons kinetic energy. The results of our calculations are illustrated in Fig.~\ref{fig:IR_colors} for carbon grains. The Spitzer colors I$_\nu (8 \mu$m)/I$_\nu (24 \mu$m) and I$_\nu (24 \mu$m)/I$_\nu (70 \mu$m) are plotted versus grain size for radiative heating by the mean Eagle Nebula radiation field, and radiative+collisional heating for a range of electron densities $n_e$. The temperature of the electrons $T_e$ is fixed to $10^6$ K. Our specific choice of $T_e$ is not critical, because the colors depend mainly on the plasma pressure, i.e. the product $n_e \times T_e$. Collisional heating has a significant impact on the infrared colors for pressures $p/k$ larger than a few $10^7\ \rm{K.cm^{-3}}$. The figure shows that both colors may be fit for pressures $p/k = 1.9\times n_e \, T_e \sim 5 \times 10^7\ \rm{K.cm^{-3}}$ and a characteristic grain size of $\sim$ 10 nm. For this plasma pressure, collisions with electrons dominate the heating of small grains with radii $< 10$ nm, while radiation is the main heating source for larger grains. To illustrate the ability of the dust model to fit the shell SED, we use a dust size distribution that combines a log-normal size distribution for very small carbon grains plus a power-law size distribution for silicates. We keep the relative fractions of dust mass in carbon grains and silicates to their interstellar values: 1/3 and 2/3, respectively. In Fig.~\ref{fig:shell_fit_collisions}, we show a fit of the ``Shell border'' SED obtained for $n_e = 30\ \rm{cm^{-3}}$ and $T_e = 10^6$ K. For this fit, the characteristic radius (i.e. the mean value of the log-normal size distribution) of the carbon VSGs is 6.5 nm. This value is somewhat smaller than the value that may be inferred from Fig.~\ref{fig:IR_colors}, because the silicates contribute to about half of the 70 $\mu$m flux. The figure shows that for a given plasma temperature the characteristic grain size is tightly constrained by the I$_\nu (8 \mu$m)/I$_\nu (24 \mu$m) ratio. It depends on the plasma temperature because this constraint is related to the stochastic heating of the smallest grains by collisions with electrons. The model also allows us to estimate the dust mass in the shell. The dust surface density is $\rm 2\times 10^{-3} \, M_\odot \, pc^{-2}$. Scaling this value by the full extent of the shell (4 pc radius), we find a total dust mass of $3\times 10^{-2}\ \rm{M_\odot}$.

The pressure inferred from the modeling of the collisional heating may be compared with independent constraints on the pressure within the Eagle nebula. This comparison raises difficulties with, but does not fully rule out, the collisional heating solution. The gas pressure inferred from Hubble observations optical line emission from the faint end of the photo-evaporation flows arising from Pillar I is $p/k \sim 10^7\ \rm{K. cm^{-3}}$ \citep[see Fig.~7b, abscissa 0 in][]{Hester1996}. This value sets an upper limit on the ambient pressure around the flows, which is lower than the pressure required for the collisional heating solution. One possible way out of this problem is that Pillar I is not embedded in the shell. The shell pressure can also be estimated from Pilbratt's blob. The blob is close to an O8.5V star known to be associated with the ionizing cluster of the Eagle Nebula (see Fig.~\ref{fig:blobowshock}). Its morphology and position on one side of the star suggests that it traces a bow shock created by a supersonic motion between the shell and the star \citep{vanBuren1990}. If this interpretation is right, it sets a constraint on the shell pressure. At the standoff distance $d_o $, i.e. the distance between the star and the edge of the blob, there is a pressure equilibrium between the wind pressure and the ambient pressure plus the ram pressure associated with the star motion. Hence, the wind pressure at the standoff distance, $ p_w =\dot{M_w} \times V_w/(4\, \pi \times d_o^2)$, is an upper limit on the ambient pressure. From the $24 \mu$m image, $d_o = 0.2\, $pc. We use the empirical relation between wind momentum and stellar luminosity \citep{Kudritzki2000}: for an O8.5V star $\dot{M_w} \times V_w \sim 2\times 10^{-7}\ \rm{M_\odot.yr^{-1}} \times 10^3\ \rm{km.s^{-1}}$. Hence, we find $p_w/k = 2\times 10^6\ \rm{K.cm^{-3}}$, a value more than one order of magnitude smaller than the pressure required for the collisional heating solution. Here, the plausible way out would be that Pilbratt's blob is not a bow-shock.

\section{The nature of the mid-IR shell}
\label{lab:nature}

\textbf{In this final part of the paper, we discuss the results from our dust modeling in the context of the Eagle Nebula massive star forming region. We have shown in the previous sections that the dust SED of the MIR shell cannot be accounted for by standard models (i.e. interstellar dust heated by UV radiation). We find two possible explanations. (1) The fraction of the dust mass in stochastically heated VSGs is much larger in the shell than in the diffuse intertsellar medium. (2) There is an additional source of heating which could be collisional heating in a high pressure plasma. Here we present two scenari that can explain either or both of these requirements. In the first one the mid-IR shell is a windblown shell, where the dust is heated by UV photons and where large grains have been ground into stochastically heated small particles. In the second scenario we investigate a more speculative hypothesis where the shell would be a supernova remnant that would be cooling through IR dust emission.}

\subsection{A wind blown shell}
\label{lab:winds}

In this first scenario, matter outflowing from dense condensations and exposed to ionizing radiation from the stellar cluster, in particular the Eagle pillars, supply the shell with a continuous inflow of gas and dust. The mechanical pressure from the stellar winds push this matter outward, but the shell persists provided that its outward expansion is compensated by continuing photo-evaporation. Since the shell is within the ionizing boundary of the nebula, the diffuse matter in the shell is fully ionized. The gas density and column density are too small to absorb all of the ionizing radiation. To quantify this scenario, we apply the empirical relation between wind momentum and stellar luminosity \citep{Kudritzki2000} to each of the O stars in the cluster. For a shell inner radius of 3 pc, we find that the winds pressure is $p_{winds}/k = 5\times 10^5\ \rm{K.cm^{-3}}$. This value is a few times larger than the radiation pressure estimated from the shell infrared brightness $p_{rad}/k \sim B_{IR}/c \sim 10^5\ \rm{K.cm^{-3}}$, where $B_{IR}$ is the mean bolometric IR brightness $\sim 0.4\ \rm{erg.cm^{-2}.s^{-1}}$ and $c$ the speed of light. The shell matter moves outward, because the wind pressure is higher than the average pressure in the interstellar medium. The expansion velocity is commensurate with the sound speed in the shell, and thus must be $\rm \sim 10\ \rm{km.s^{-1}}$. Since the shell is a few parsecs wide, the shell matter needs to be renewed over a timescale of a few $10^5\ \rm{yr}$ by on-going photo-evaporation.

In the Eagle Nebula, the pressure from stellar winds is too low to account for the shell colors with collisional excitation \textbf{(see section \ref{lab:collheating} for details)}. The mechanical power from the winds is also too small to contribute to the IR luminosity from the shell. For a wind velocity of $2500\ \rm{km.s^{-1}}$ \citep{Kudritzki2000}, the mechanical energy injection is $\sim 2500\ \rm{L_\odot}$, a factor 20 smaller than the shell luminosity $\sim 5 \times 10^4\ \rm{L_\odot}$ as estimated from the shell brightness $B_{IR}$ and its angular diameter (14'). Unlike what \citet{Everett2010} advocated for N49, in M16 the shell IR emission cannot be powered by the stellar winds, and does not represent a major cooling channel that impacts the dynamical evolution of a wind-blown shell.

\textbf{The shell must originates from the only available source of dust, i.e., evaporating dense gas condensations within the ionization boundary of the Nebula. The difficulty in being certain that this is the right interpretation comes from interstellar dust (see sections \ref{lab:uvheating} and \ref{lab:collheating}). Indeed, our dust modeling in section \ref{lab:uvheating} shows that the shell SED cannot be fit with the standard interstellar dust size distribution. The fits shown in Figure \ref{fig:shell_sed_plot_fit} and \ref{fig:otherfit} illustrate the uncertainty of the modeling. It is beyond the scope of this paper to explore in a systematic way the full range of possible solutions, but we are confident that any fit will involve shattering of dust grains to nanometric sizes.}

As a consequence of such an interpretation for the Eagle Nebula shell, we conclude that massive star forming regions have a major impact on carbon dust. \citet{Galliano2003} reached a similar conclusion in their modeling of the infrared SED of the dwarf, star forming, galaxy NGC~1569. Observations of the ionized gas kinematics do provide evidence for supersonic velocities in the immediate environment of pillars in star forming regions \citep{Westmoquette2009}. Hence, the grinding of the carbon dust could be the result of grain shattering in grain-grain collisions within shocks driven by the dynamical interaction between the stellar winds and the shell. Theoretical modeling of the dust dynamics in shocks suggest that this is a plausible hypothesis \citep{Jones2004}. \citet{Guillet2009} have quantified dust processing by the passage of J-shocks of a few 10 $\rm{km.s^{-1}}$. They find that the mass fraction in the largest grains is reduced to the profit of the smallest, as a result of grain shattering and dust vaporization. 

\subsection{A supernova remnant}
\label{lab:SNR}

\textbf{Alternatively, we keep the usual distribution of dust grain sizes, but look for another source of pressure: a supernova remnant. This is not unexpected for a 3-Myr old nebula with very massive stars \citep[$M_\star \sim 80 M_\odot$][]{Hillenbrand1993}. If so, we would be witnessing a specific time in the evolution of the remnant where the plasma pressure and temperature would be such that the remnant cools through dust emission. This scenario relates directly to the fit of the shell SED quantified in section~\ref{lab:collheating}.}

The infrared dust emission from fast shocks driven by supernovae has been quantified in several theoretical papers \citep[e.g.][]{Draine1981, Dwek1996}. Overall, dust is found to be a significant but not dominant coolant of shocked plasma due to dust destruction. For a dust to hydrogen mass ratio of 1\% and a Solar metallicity, dust cooling is larger than atomic cooling for temperatures $> 5 \times 10^5$ K, but, for temperatures T larger than $\sim 10^6$ K, the dust destruction timescale by sputtering is smaller than the gas cooling time \citep{Smith1996, Guillard2009}. This framework has been used to interpret observations of young remnants starting from the first infrared detections of supernovae with the IRAS survey \citep{Dwek1987}. We propose here a distinct idea, where the shell infrared emission seen towards the Eagle Nebula would be related to the late evolution of a remnant. 

For the model shown in Fig.~\ref{fig:shell_fit_collisions}, 1/3 of the shell infrared emission is powered by grain collisions with electrons and contributes to the plasma cooling. The remaining 2/3 is provided by radiative heating of the dust. Assuming that the dust infrared emission is the dominant gas cooling channel, the isobaric cooling time of the infrared emitting plasma is $ t_{cool} = {{5}\over{2}}\times 2.3 \times k\, T_e / (\Gamma \times m_p \times x_d)$ where $ \Gamma$ is the collisional heating rate per unit dust mass, $m_p $ the proton mass and $x_d$ the dust to hydrogen mass ratio. With the $ \Gamma$ value derived from the fit in Fig.~\ref{fig:shell_fit_collisions}, we find $t_{cool} = 1500\times (x_d/0.01)^{-1}$ yr. The dust-to-hydrogen mass ratio $x_d$ is not constrained by the modeling. This factor may well be smaller than the reference value of 1\% due to dust destruction by sputtering. The SED fit also allows us to estimate the plasma column density and thereby the internal energy $U$ of the infrared emitting plasma. The model gives $N_H = 8\times 10^{18} \times (x_d/0.01)^{-1}\ \rm{H.cm^{-2}}$. From there we find $U = 2\times 10^{48} \times (x_d/0.01)^{-1}\ \rm{erg}$. This value is a small fraction of the expansion energy associated with a typical supernova explosion ($\sim 10^{51}\ \rm{erg}$). Within our remnant hypothesis, this large difference indicates that the cooling time is short and that only a small fraction of the shocked plasma is contributing to the infrared emission. One possibility to account for this fact would be that we are observing the late evolution of the remnant when the low density hot plasma heated to high temperatures early in the expansion of the remnant is cooling through turbulent mixing with photo-ionized gas \citep{Begelman1990}. This plasma would have a long intrinsic cooling timescale, because its dust would have been destroyed early in the evolution of the remnant. For a pressure of $p/k = 5\times 10^7\ \rm{K.cm^{-3}}$, the cooling timescale through atomic processes of a dust-free plasma at a temperature of $10^7$ K is $ 2 \times 10^6$ yr. 

\textbf{This interpretation will need to be tested against additional observations. The absence of bright diffuse emission in the Chandra X-ray images \citep{Linsky2007} can possibly be accounted for. For instance, the hot plasma may be too tenuous to be seen in emission, while the X-ray emission from the turbulent mixing layers would be soft and thus heavily attenuated by foreground gas. We re-analyzed the Chandra ACIS-I observations of M16 \citep{Linsky2007} to search for a {\it faint} background emission. After removal of point sources, we do find residual X-ray emission over the SW section of the mid-IR shell where the foreground extinction is the lowest. The emission spectrum fit gives $kT$ in the range $0.6-2\ \rm{keV}$ and a foreground column density within $2.4$−$5.4 \times 10^{22}\ \rm{H.cm^{−2}}$. The absorption corrected X-ray brightness is $1.3 \times 10^{−3}\ \rm{erg.s^{−1}.cm^{−2}.sr^{−1}}$. If this emission arises from the mid-IR shell (i.e. from a sightline length $\sim 10$ pc), we derive a plasma pressure $p/k \sim 10^8\ \rm{K.cm^{−3}}$. This result does not allow us to conclude that the X-ray emission arises from a supernova remnant, but, if it does, the X-ray emission is consistent with the dust being collisionally excited in a high pressure plasma. In this case, if the X-ray emission fills the mid-IR cavity, the shell X-ray luminosity would be $\sim 10^{33}\ \rm{erg.s^{−1}}$. This is on the low side for an SNR: for comparison, the W28 SNR, which is interacting with a molecular cloud, has a total X-ray luminosity $L_X \sim 6 \times 10^{34} {\rm erg\,s}^{-1}$ \citep{Rho2002}. However, our value of $L_X$ for the putative M16 SNR is a lower limit, since it does not take into account the soft X-ray emission from cooler gas that is more heavily absorbed. Further X-ray observations are planned to clarify this point. MIR spectroscopic maps of M16 with Spitzer, covering a wide range of emission features and ionization energies, will provide an additional test to be investigated.}

\section{Conclusions}
\label{lab:ccl}

\begin{itemize}
\item We present new IR images of the Eagle Nebula from the MIPSGAL survey that reveal the well-known illuminated clouds of dust and gas. The MIPS 24 $\mu$m observations shows the same inner shell-like feature as mid-infrared observations from ISO or MSX. It is significantly brighter than the PDRs. Relative to these previous observations, the MIPSGAL survey has the advantage to also probe the far infrared emission of the dust. The structure of the nebula as seen in the MIPS 70 $\mu$m observations is close to that of the shorter wavelengths as seen in the GLIMPSE survey (from 3 to 8 $\mu$m): the cloud surface is significantly brighter than the inner shell.

\item Thanks to the MIPS 24 and MIPS 70 $\mu$m observations, we are able to give constraints on the temperature of the grains emitting in the FIR range and the required interstellar radiation field intensity to heat them up to these temperatures with our dust model. The dust temperature varies from $\sim$35 K in the PDRs to $\sim$70 K in the shell. The required intensity of the ISRF within the PDRs is about an order of magnitude lower than that provided by the star cluster NGC6611. The shell of hot dust, however, requires an ISRF intensity about a factor of a few higher than that provided by the cluster. 

\item Combining all the IR observations at our disposal into SEDs that sample the whole nebula with our dust model, we fit the observations to constrain both the radiation field intensity and the dust size distribution. In the PDRs, we confirm the required ISRF intensity is about a few tenth of that provided by NGC6611. The dust size distribution is dominated by BGs even though all the dust components are present with abundance a factor of a few, at most, different from those of the DHGL medium. In the shell, we also confirm the required ISRF intensity is a factor of a few larger than that of NGC6611. The PAHs are absent and the VSGs are more abundant, up to a factor 10, than in the DHGL medium.

\item Extinction and the dispersion of the stars across the nebula can account for the lower ISRF intensity required for the PDRs. On the contrary, an additional source of heating is required for the shell. Neither the spatial variations of the ISRF intensity nor the Lyman alpha photons contribution can account for the discrepancy between required and provided UV heating of the dust. Exact positions of the stars reveal that Pilbratt's blob is only 0.25 pc from an O8.5V star and may thus be a bow shock.

\item We then invoke gas-grain collisions as an extra source of heating. Our modeling leads to a fit of the shell SED that requires a pressure of a few $10^7\ \rm{K.cm^{-3}}$. Such a pressure is at least a factor of a few larger than that inferred either from optical observations at the end of the photo-evaporation flow arising from Pillar or from Pilbratt's blob bow shock nature.

\item \textbf{We finally discuss two interpretations of the mid-IR shell in the general context of a massive star forming region. In a first scenario, we propose that the shell is wind blown by the stars. We find that the star cluster does not provide enough mechanical energy via stellar winds to power the shell emission. Therefore the shell is explained by a modified dust grain size distribution (large carbon grains shattered to nanometric sizes) with heating only due to UV emission. The implication is then that massive star forming regions like M16 have a major impact on their dust size distribution: this can be checked on other similar regions. Alternatively, we propose a second scenario, in which the shell is heated by the hidden remnant of a supernova from a very massive progenitor, and for which the dust provides a fast cooling. The implication is then that our observations occur during a short-lived, late stage of evolution of the remnant: this can be checked with new X-ray observations.}


\end{itemize}
 
The Eagle Nebula IR emission morphology is similar to that of many other star forming regions observed within the GLIMPSE and MIPSGAL surveys \citep{Churchwell2006, Carey2009}. For the first time, it is quantitatively discussed in terms of dust modeling. The work we present would need to be extended to other SFRs with IR morphology similar to that of M16 to ascertain whether the interpretation would be challenged by the same problem in accounting for the dust temperature. Moreover, future analysis of additional observations (mid-to-far IR spectral mapping from Spitzer/IRS and MIPS-SED, near-IR narrow band imaging from CFHT/WIRCam) of the Eagle Nebula will provide us with more constraints on the physical conditions and dust properties in M16's inner shell.

This work is based in part on observations made with the Spitzer Space Telescope, which is operated by the Jet Propulsion Laboratory, California Institute of Technology under a contract with NASA. Support for this work was provided by NASA through an award issued by JPL/Caltech.


\bibliographystyle{aa} 

\end{document}